\documentclass[%
 reprint,
superscriptaddress,
 amsmath,amssymb,
 aps,showpacs
]{revtex4-2}
\usepackage{amssymb}
\usepackage{lineno}
\usepackage{hyperref}
\usepackage{siunitx}
\usepackage{wasysym}
\usepackage{gensymb}
\usepackage{tabularx}
\usepackage{braket}
\usepackage{physics}
\usepackage{url}
\usepackage{todonotes}
\usepackage{comment}
\usepackage{xcolor}
\usepackage{booktabs,multirow}%

\usepackage{paracol}
\usepackage{wrapfig}
\usepackage{subcaption}
\usepackage{mwe}
\usepackage{graphicx}
\usepackage{dcolumn}
\usepackage{bm}

\begin{document}

\preprint{APS/123-QED}

\title{Entanglement swapping systems toward a quantum internet}

\thanks{DO NOT DISTRIBUTE/CONFIDENTIAL/DRAFT }%

\author{Samantha I. Davis}
\author{Raju Valivarthi}
\affiliation{Division of Physics, Mathematics and Astronomy, California Institute of Technology, Pasadena, CA 91125, USA}
\affiliation{Alliance for Quantum Technologies (AQT/INQNET), California Institute of Technology, Pasadena, CA 91125, USA}
\author{Andrew Cameron}
\affiliation{Fermi National Accelerator Laboratory, Batavia, IL 60510, USA}
\author{Cristi\'{a}n Pe\~{n}a}
\affiliation{Fermi National Accelerator Laboratory, Batavia, IL 60510, USA}
\author{Si Xie}
\affiliation{Fermi National Accelerator Laboratory, Batavia, IL 60510, USA}
\affiliation{Division of Physics, Mathematics and Astronomy, California Institute of Technology, Pasadena, CA 91125, USA}
\affiliation{Alliance for Quantum Technologies (AQT/INQNET), California Institute of Technology, Pasadena, CA 91125, USA}
\author{Lautaro Narv\'{a}ez}
\affiliation{Division of Physics, Mathematics and Astronomy, California Institute of Technology, Pasadena, CA 91125, USA}
\author{Nikolai~Lauk}\altaffiliation{Present Address: Photonic Inc., Coquitlam, BC V3K 6T1, Canada}

\author{Chang~Li}\altaffiliation{Present Address: Joint Quantum Institute, University of Maryland, College Park, MD 20742, USA}
\author{Kelsie~Taylor}
\altaffiliation{Present Address: Weinberg Institute for Theoretical Physics, University of Texas at Austin, Austin, TX 78712, USA}
\author{Rahaf~Youssef$^{1}$}
\altaffiliation{Present Address: University of Chicago, Chicago, IL 60637, USA}
\affiliation{Alliance for Quantum Technologies (AQT/INQNET), California Institute of Technology, Pasadena, CA 91125, USA}
\author{Christina~Wang $^{1}$}
\affiliation{Fermi National Accelerator Laboratory, Batavia, IL 60510, USA}

\affiliation{Alliance for Quantum Technologies (AQT/INQNET), California Institute of Technology, Pasadena, CA 91125, USA}

\author{Keshav Kapoor}
\affiliation{Fermi National Accelerator Laboratory, Batavia, IL 60510, USA}\altaffiliation{Present Address: University of Illinois, Champgaign, IL 61820, USA}
 

\author{Boris Korzh}
\affiliation{Jet Propulsion Laboratory, California Institute of Technology, Pasadena, CA 91109, USA}
\affiliation{Alliance for Quantum Technologies (AQT/INQNET), California Institute of Technology, Pasadena, CA 91125, USA}
\author{Neil Sinclair}
\affiliation{Division of Physics, Mathematics and Astronomy, California Institute of Technology, Pasadena, CA 91125, USA}
\affiliation{Alliance for Quantum Technologies (AQT/INQNET), California Institute of Technology, Pasadena, CA 91125, USA}
\affiliation{John A. Paulson School of Engineering and Applied Sciences, Harvard University, Cambridge, MA 02138, USA}

\author{Matthew Shaw}
\affiliation{Jet Propulsion Laboratory, California Institute of Technology, Pasadena, CA 91109, USA}
\author{Panagiotis Spentzouris}
\affiliation{Fermi National Accelerator Laboratory, Batavia, IL 60510, USA}

\author{Maria Spiropulu}
\affiliation{Division of Physics, Mathematics and Astronomy, California Institute of Technology, Pasadena, CA 91125, USA}
\affiliation{Alliance for Quantum Technologies (AQT/INQNET), California Institute of Technology, Pasadena, CA 91125, USA}

\date{\today}

\begin{abstract}

We demonstrate conditional entanglement swapping, i.e. teleportation of entanglement, between time-bin qubits at the telecommunication wavelength of 1536.4 nm with high fidelity of 87\%. Our system is deployable, utilizing modular, off-the-shelf, fiber-coupled, and electrically controlled components such as electro-optic modulators. It leverages the precise timing resolution of superconducting nanowire detectors, which are controlled and read out via a custom developed graphical user interface. The swapping process is described, interpreted, and guided using  characteristic function-based analytical modeling  that accounts for realistic imperfections. Our system supports quantum networking protocols, including source-independent quantum key distribution, with an estimated secret key rate of approximately 0.5 bits per sifted bit.
\end{abstract}

\maketitle

\section{Introduction}\label{sec:intro}

A quantum internet, a large-scale quantum network, aims to distribute entangled qubits over long distances and between disparate quantum hardware~\cite{kimble2008quantum,simon2017towards,wehner2018quantum}. For metropolitan-distance networks, qubits are encoded into photons, with fiber optics as the preferred medium for transfer~\cite{sangouard2011quantum,dias2017quantum,lucamarini2018overcoming,bhaskar2020experimental}. To mitigate loss, photons at telecommunication wavelengths, such as the 1550 nm C-band, are used~\cite{Valivarthi2016,Sun2016,Takesue2009,thomas2024quantum}. Since loss scales exponentially with fiber length, multiplexing, quantum repeaters, or a combination of both can ensure qubits traverse a channel. These techniques also improve the generation rate of single and entangled photons created using probabilistic processes like spontaneous parametric down conversion (SPDC)~\cite{migdall, sinclair2014spectral, saikatmultiplex}.

Entanglement swapping, where a Bell-state measurement (BSM) entangles qubits that have never interacted, is crucial for entangling remote qubits and enabling quantum repeaters~\cite{Azuma2023repeaterreview}. Since the first demonstration of post-selective and conditional entanglement swapping of photons~\cite{Pan1998firstswapping}, numerous follow-up experiments have focused on quantum communications~\cite{halder2007entangling,Sun2017_100km_vis73,Samara2021IntegratedSwapping,Liu2022alloptical}. Entanglement swapping also has applications in quantum computing~\cite{Li2014computation}, quantum sensing~\cite{gottesman2012telescope}, and fundamental tests of quantum mechanics~\cite{Branciard2010}. Various renditions of photonic entanglement swapping have been demonstrated, including using qubits encoded into different degrees of freedom~\cite{Kaltenbaek2009polarizationswapping} or derived from different sources~\cite{Basset2019QuantumdotSwappingFid58}.

Time-bin encoding is advantageous for quantum networks because each logical state is encoded into the same degrees of freedom except time. This avoids unintended mode-dependent transformations and phase shifts. It also allows simple interfacing of quantum hardware, such as atomic memories or optical frequency converters, which are generally not compatible with multiple modes. Time-bin encoding provides access to high-dimensional states, i.e., qudits, that encode more information than qubits, benefiting quantum communications~\cite{Cerf2002_high_d_cryptography, erhard2018twisted} and computing~\cite{Campbell2012magicstatedist}.

Thus far, entanglement swapping of photonic time-bin qubits has yielded states with an average fidelity up to 83\% \cite{Reidmatten2005,halder2007entangling,Sun2017_100km_vis73}. Fidelity $F=\bra{\psi} \rho \ket{\psi}$ of the swapped state $\rho$ with respect to the state $\ket{\psi}$ \cite{nielsen2001quantum} is an important figure of merit to optimize for in quantum networks and information applications in general. Non-classical states manifest at a fidelity greater than 50\% with respect to a target Bell state. Clauser-Horne-Shimony-Holt inequality violations, which are signatures of non-locality and benefit fundamental tests \cite{zukowski1993event,Peres2000DelayedChoiceSwapping,ma2012experimental}, occur for a fidelity greater than 78\% \cite{CHSH}. The Ekert protocol for entanglement-based quantum key distribution and source-independent quantum key distribution based on swapping require a fidelity greater than 89\% \cite{ekert1991quantum, ma2007quantum}. Distributed quantum computing likely requires fidelity greater than 99.999\% \cite{nielsen2001quantum,ladd2010quantum,gisin2007quantum,pirandola_advances_2015,briegel1998quantum, rudolph2017optimistic}. The unavoidable presence of loss in networks further supports the pursuit of high fidelity to reduce the impact of photon counting statistics.

In this work, we demonstrate conditional entanglement swapping between two degenerate time-bin entangled photonic qubits at the telecommunication wavelength of 1536.4 nm with an average fidelity greater than 87\%. This fidelity allows demonstration of source-independent quantum key distribution, a scheme that assumes qubits may be generated by an adversary, with an estimated secret key rate of approximately 0.5 bits per sifted bit. The qubits are created with modular, off-the-shelf, fiber-coupled, and electrically controlled components, facilitating setup reproduction and deployment in networks. Specifically, we generate an entangled state using SPDC in nonlinear waveguides pumped with two visible-wavelength pulses separated by 346 ps. Electro-optic modulators carve two pulses from continuous-wave laser light at 1536.4 nm, which are upconverted using another nonlinear waveguide.

Projection onto the Bell state $\ket{\Psi^-}$ and subsequent measurement of the swapped state $\ket{\Phi^+}$ (up to a known phase) with Michaelson interferometers is facilitated by superconducting nanowire single photon detectors (SNSPDs) that resolve the 346 ps bin separation. Our experiment is facilitated with semi-autonomous control, monitoring, and synchronization, with all data collected using scalable software and hardware. The system yielded swapping rates of 0.01 Hz at a clock rate of 200 MHz and was run remotely over several days. The experiment was interpreted and guided using characteristic function-based analytical modeling based on realistic imperfections. Based on our modeling, we identify straightforward improvements, such as improved packaging and integration to reduce loss, and reduction of the bin separation to the ps-scales, which is compatible with state-of-the-art modulators \cite{prash2021breaking, zhu2021integrated} and SNSPDs \cite{korzh_demonstration_2020,Mueller:2023uoj}, to improve the swapping rates to approximately Hz without compromising fidelity. Finally, our specific choice of wavelength is compatible with quantum emitters, memories, and transducers using erbium-doped crystals \cite{zhong2019emerging}. The demonstration extends our previous work using quantum teleportation systems \cite{valivarthi2020teleportation} toward a workable quantum internet envisioned by the U.S. Department of Energy to link the U.S. National Laboratories\cite{QI-blueprint}.

\section{Setup}\label{sec:setup}
\begin{figure*}[htbp!]
  \centering
   \includegraphics[width=\textwidth]{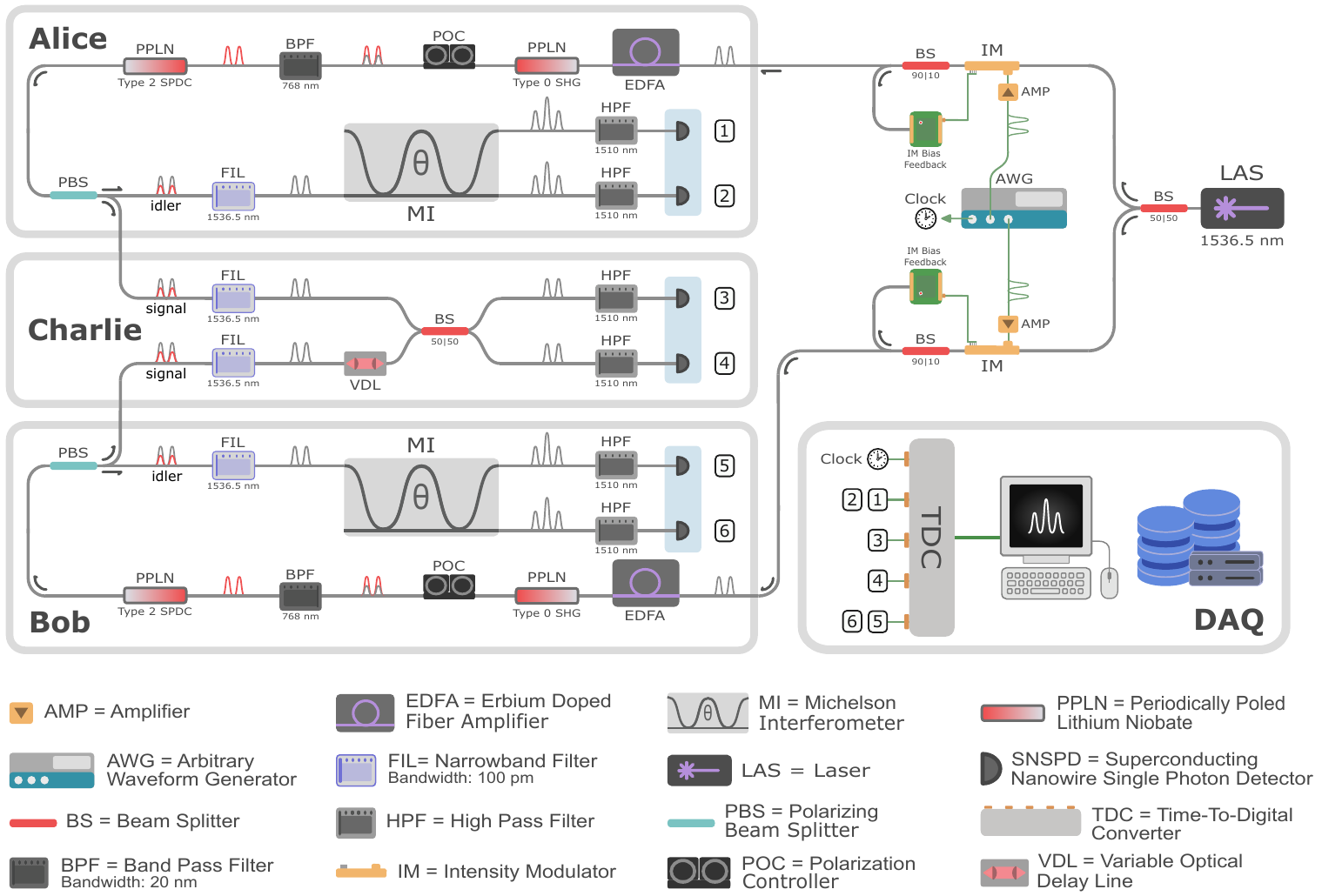}
   \caption{Schematic diagram of the entanglement swapping system consisting of Alice, Bob, Charlie, and the data acquisition (DAQ) subsystems. All components are labelled in the legend. Single mode fibers and electronic cables are indicated in gray and green, respectively. The detection signals generated by the SNSPDs are labeled 1-6 and sent to the TDC, with 1-2 and 5-6 time multiplexed. The clock generated by the AWG is labeled and sent to the start channel of the TDC.
   }
   \label{fig:setup}
 \end{figure*}

The setup for entanglement swapping is shown in Fig.~\ref{fig:setup}.
We demonstrate a swapping protocol in which a qubit of an entangled photon pair (from Alice) is interfered with a qubit of another entangled photon pair (from Bob) and then measured in the Bell state $\ket{\Psi^-}$ (at Charlie). 
As a result, the remaining photons at Alice and Bob are projected onto a Bell state $\ket{\Phi^+}$, which is defined with respect to a pre-determined phase offset.
All qubit measurements are performed with a custom developed data acquisition (DAQ) system. 
The Alice, Bob, Charlie, and DAQ subsystems are detailed in the following subsections.


\subsection{Alice and Bob: entangled photon pair generation }\label{subsec:EPS}
To generate the entangled photon pairs, light from a fiber-coupled continuous wave (CW) laser at a telecom wavelength of 1536.4~nm is split into two paths by a 50:50 polarization-maintaining beamsplitter (BS).
In each path, the light is coupled into a lithium niobate intensity modulator (IM) driven by an arbitrary waveform generator (AWG).
The AWG generates a pair of pulses separated by 346~ps at a repetition rate of 200~MHz.
Each pulse has a full-width-at-half-maximum (FWHM) of approximately 65~ps.
The pulses from the AWG are amplified by a 30~dB high-bandwidth amplifier and are injected into the radio-frequency (RF) input of the IM, resulting in optical pulses with an extinction ratio of at least 20~dB.
A 90:10 BS at the output of the IM is used to perform feedback on the DC-bias port of the IM, which ensures a constant extinction ratio throughout the experiment.
The optical pulses from the 90\% ports of the BS in each path are sent to the Alice and Bob nodes.

At each node, the optical pulses are amplified with an erbium-doped fiber amplifier (EDFA) and up-converted to 768.2~nm by second harmonic generation (SHG) with a periodically poled lithium niobate (PPLN) waveguide.
Residual pump light at 1536.4~nm is removed by a 768~nm bandpass filter with an extinction ratio of $\geq 80$ dB.
The up-converted pulse pair is used to pump a type-II SPDC with a second PPLN waveguide, which produces a pair of photons at 1536.4~nm in a Bell-state state, $\ket{\Phi^+} = (\ket{ee}+\ket{\ell\ell})/\sqrt{2}$.
The members of the entangled photon pair are produced in ``signal" and ``idler" modes with orthogonal polarizations $\ket{H}$ and $\ket{V}$ and are hereafter referred to as ``signal" and ``idler" photons, respectively.
The signal and idler photons are separated with a polarizing beam splitter (PBS) and spectrally filtered with tunable narrowband optical filters. 
We select a bandwidth of 12.7~GHz to optimize for the trade-off in spectral purity and Bell pair generation rates. 
The signal photon is distributed to Charlie for the Hong-Ou-Mandel or Bell State measurements, and the idler photon is sent to an unbalanced Michelson interferometer (MI) with a delay of 346~ps between the long path and the short path.
The path difference matches the time between early and late time-bins and enables pulse overlap required for characterizing entanglement.
When two time-bins enter an MI, three time-bins emerge corresponding to all combinations of short and long path trajectories of each input time-bin.

All single photon detections are performed with SNSPDs.
A total of six SNSPDs, two each for Alice, Bob, and Charlie, are installed in a rack-mount cryogenic system with continuous operation at a temperature of 2.5~K.
The SNSPDs have detection efficiencies of 89-93\%, dark count rates of 60-135~Hz, timing resolution (jitter) of 38-59~ps, and dead times $\leq 30$~ns.
A photon detection at the SNSPD produces an RF pulse, which is sent to the DAQ subsystem described in Sec. \ref{subsec:DAQ}.
At the input of each detector, a high pass filter is used to remove any residual 768~nm light from the second harmonic generation process.

The MI and SNSPDs are used to project a photon onto the $\ket{e}$, $(\ket{e}+e^{i\theta}\ket{\ell})/\sqrt{2}$, or $\ket{\ell}$ states by detection in the first, second, or third time-bins, respectively, at one of the outputs (see Fig~\ref{fig:setup}). 
Detection at the other output corresponds to projections onto the same states but with $\theta +\pi$.
The phase $\theta$ of a MI is set by the voltage applied to its phase shifter. 
For the entanglement visibility measurements, which include the characterization of entangled photon pair sources and teleportation of entanglement, the phase of Alice's MI is swept and the coincidence events of photons in the outputs of Alice's and Bob's MI are accumulated.
We measure the coincidences in all four pairings of Alice and Bob's outputs to maximize the coincidence rates in the experiment.

\subsection{Charlie: Bell-state measurement}\label{subsec:BSM}
At Charlie, the signal photons from Alice and Bob are interfered in a 50:50 polarization maintaining beamsplitter (BS) after spectral filtering. 
A variable optical delay line (VDL) at one input of the BS is used to optimize the temporal indistinguishability of the interfering photons, such that the photons arrive to the inputs of the BS at the same time.
Alice and Bob's signal photons are projected onto the  $\ket{\Psi^{-}} = (\ket{e\ell}-\ket{\ell e})/\sqrt{2}$ Bell state by detection of coincidence events in the first time-bin of one BS output and the second time-bin of the other BS output. 
Conditioned on a successful Bell-state measurement outcome, Alice and Bob's idler photons are projected onto the $\ket{\Psi^{-}}$ state.

\subsection{DAQ: Data acquisition and analysis}\label{subsec:DAQ}

Our DAQ subsystem is an extension of the control and data acquisition systems detailed in~\cite{valivarthi2020teleportation}.
The RF pulses from the SNSPDs are sent to a time-to-digital converter (TDC) with a fixed voltage threshold, i.e. ``time-tagger", to obtain a time-tag for the time-of-arrival of each pulse relative to a clock signal. 
The time-tagger has five input channels, one of which is used for a 10 MHz clock signal reference from the AWG.
Of the remaining four channels, two are used for the outputs of Charlie's SNSPDs, one is used for the outputs of Alice's SNSPDs, and one is used for the outputs of Bob's SNSPDs.
The outputs for each pair of detectors at Alice and Bob are electronically combined with a relative time delay introduced by an RF delay line to enable signals from the pair to be read out with a single time-tagger channel. 
The time-tagger is interfaced with a custom graphical user interface (GUI) to process the time-tags, perform the coincidence logic, store measurement outcomes in a customized database, and visualize photon statistics in real-time (see Appendix~\ref{sec:gui}).
The database forms the backbone of a centralized classical processing unit that is responsible for the monitoring of critical network parameters, remote control, active-feedback and stabilization of experimental components, acquisition and management of large volumes of time-tagged signals, and global synchronization across multiple nodes.
The DAQ subsystem has been upgraded to support GHz teleportation rates, multinode entanglement distribution, and picosecond synchronization for metropolitan scale quantum network testbeds~\cite{valivarthi2022picosecond,kapoor2023picosecond}. 

\section{Experimental Results} \label{sec:results}

\begin{figure*}[htbp!]
   \centering
   \includegraphics[width=\textwidth]{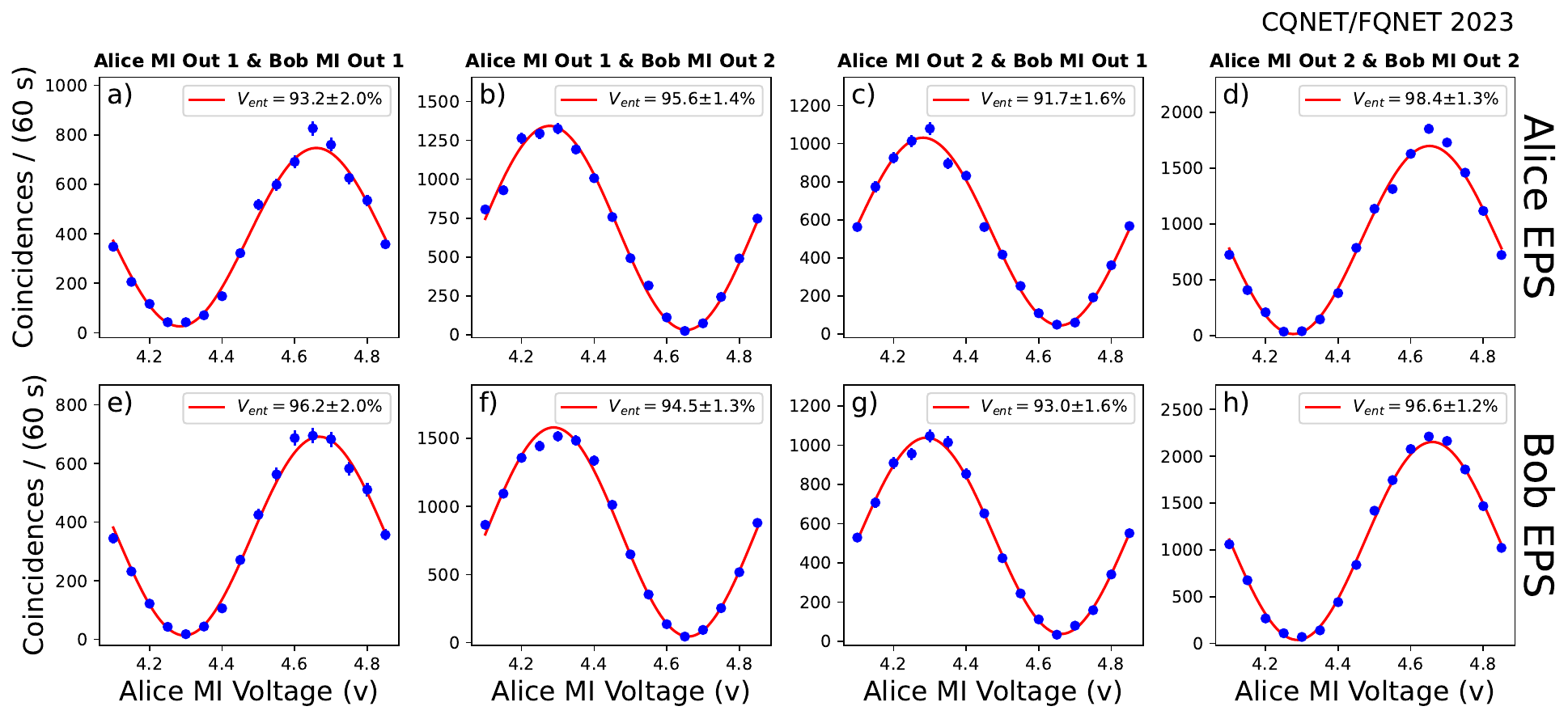}
   \caption{Entanglement visibility of photon pairs produced by Alice's and Bob's entangled photon pair source (EPS).  The coincidence rates for each pairing of an output port of Alice's MI and Bob's MI are shown for Alice's EPS a)-d) and Bob's EPS d)-h). The entanglement visibilities are obtained from a sinusoidal fit (see main text for details), with uncertainties in all measurements calculated assuming Poisson statistics.
   }
\label{fig:entanglement_visibility}
\end{figure*}

\subsection{Entanglement visibility}\label{entvisib}

The entanglement swapping protocol requires high-fidelity entanglement sources, which we realize with an SPDC process using a bulk optical nonlinearity. 
The output state of an SPDC process can be described by the two-mode squeezed state (TMSV), 
  \begin{align}
     \ket{\text{TMSV}} = \sum_{n=0}^{\infty}(-1)^n \sqrt{\frac{\mu^n}{(1+\mu)^{n+1}}}\ket{n,n}, \label{eq:TMSV}
 \end{align}
 where $\mu$ is the mean number of photon pairs, and $\ket{n,m}\equiv \ket{n}\otimes\ket{m}$ is the product state of $n$ photons in the signal mode and $m$ photons in the idler mode. 
 After the narrowband filters, the output state of the pair source is described by~$\ket{\text{pair}} = \ket{\text{TMSV}}_e \otimes \ket{\text{TMSV}}_\ell$, where $\ket{\text{TMSV}}_{e(\ell)}$ is a TMSV in the early (late) time-bin.  For low mean photon numbers, $\ket{\text{pair}}$ approximates a Bell state conditioned on the presence of at least one photon,
 \begin{align}
     \ket{\text{pair}}\approx \sqrt{1-2\mu}\ket{0} + \sqrt{2\mu}\ket{\Phi^+} + \mathcal{O}(\mu^2), \quad \mu \ll 1,\label{eq:pair}
 \end{align}
 neglecting loss. Due to multiphoton effects arising from $\mathcal{O}(\mu^2)$ contributions, there is a trade-off in the quality of entanglement and the pair production rate $\propto \mu$.
We optimize for this trade-off by operating the sources at Alice and Bob with a mean photon number per time-bin of $\mu_A = 2.5\times 10^{-3}$ and $\mu_B = 2.0\times 10^{-3}$, respectively, at a repetition rate of 200 MHz.

To evaluate the entanglement sources, we measure the entanglement visibilities of the photon pairs produced by each pair source with a modification of the setup in Fig. \ref{fig:setup}. After the narrowband filters, the signal and idler modes of a pair source are directed to Alice's MI and Bob's MI, respectively. We vary the phase of Alice's MI and measure the coincidences of the signal and idler modes in a phase basis by accumulating coincidence events in the central time-bin of both Alice and Bob's MI.  We acquire data for all four combinations of Alice and Bob MI output ports, which are used in the entanglement swapping visibility measurements. The results are shown in Fig. \ref{fig:entanglement_visibility}. 
The coincidence rates are fitted proportional to $1+V_{\text{ent}}\cos{(2\omega v + \phi_0)}$,  where the entanglement visibility is $V_{\text{ent}} = (C_{\text{max}} - C_{\text{min}})/(C_{\text{max}}+C_{\text{min}})$, with $C_{\text{max}(\text{min})}$ denoting the maximimum (minimum) rate of coincidence events, $\omega$ and $\phi_0$ are unconstrained constants, and $v$ is the voltage applied to Alice's MI.  We obtain average entanglement visibilities across all output port combinations of $\langle V_{\text{ent}}\rangle = 94.7\pm 1.6\%$ for Alice's source and $\langle V_{\text{ent}}\rangle =95.1\pm1.6\%$ for Bob's source. The deviations from unity are attributed to mulitphoton effects and interferometric imperfections. Imbalances in the MIs due to imperfect transmittances and internal path efficiences can give rise to a dependence of the entanglement visiblity on the combination of output ports that is used \cite{Mueller:2023uoj}. Nevertheless, these visibilities exceed the locality bound of $1/\sqrt{2}$, and correspond to average state fidelities with respect to $\ket{\Phi^+}$ of $\langle {F}_{\text{ent}}\rangle =  96.0\pm1.2\%$ for Alice and $\langle {F}_{\text{ent}}\rangle = 96.3\pm1.2\%$ for Bob, where $F_{\text{ent}} = (3V_{\text{ent}}+1)/4$.

\begin{figure*}[htbp!]
  \centering
    \includegraphics[width=\textwidth]{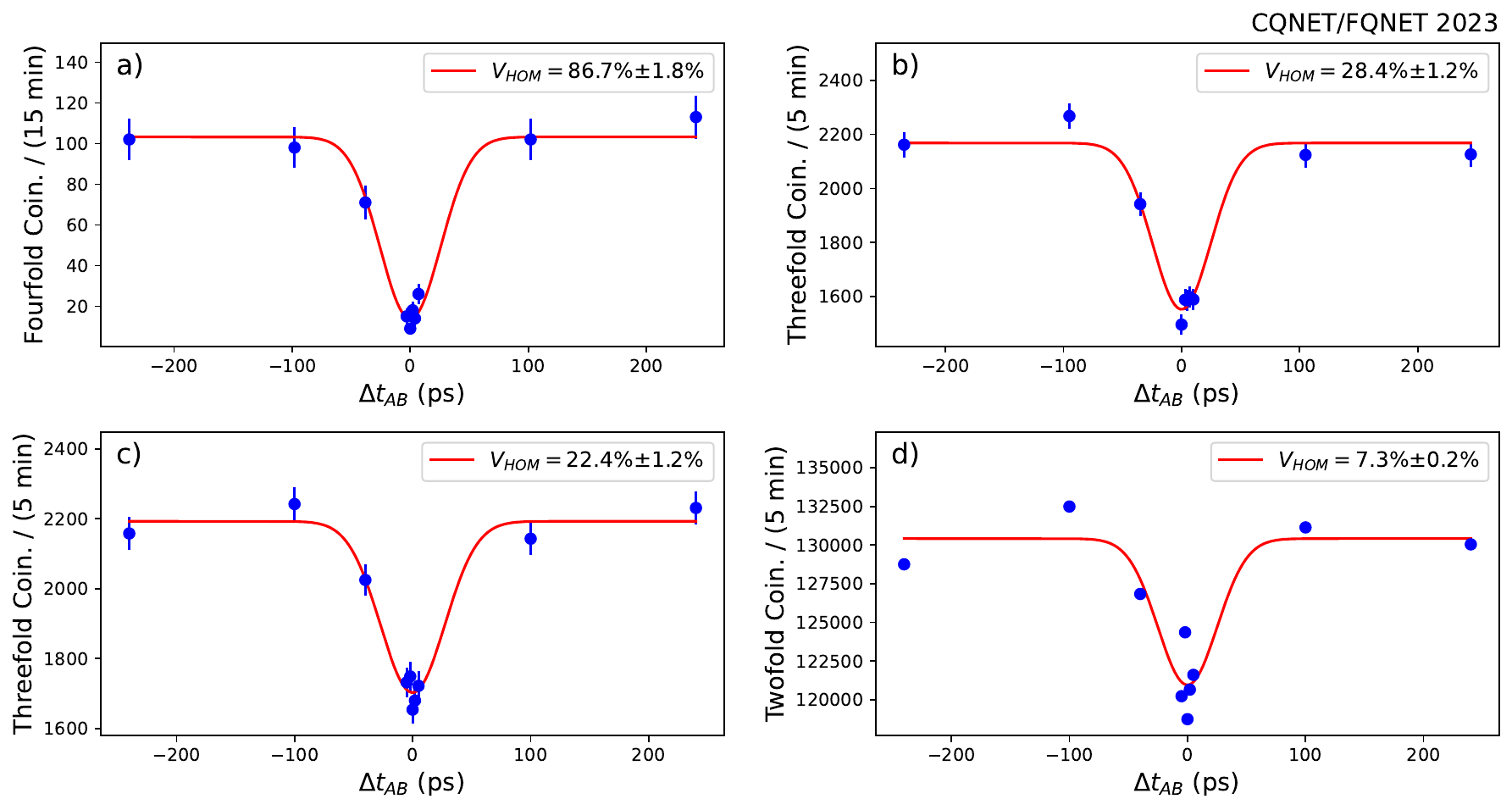}
  \caption{Hong-Ou-Mandel (HOM) interference.  a) Fourfold coincidence rates, b) threefold coincidence rates conditioned on Bob's idler photon, c) threefold coincidence rates conditioned on Alice's idler photon, and d) twofold coincidence rates measured as a function of the relative time-delay ($\Delta t_{AB}$) between Alice and Bob's signal photons.}
  \label{fig:hom}
\end{figure*}

\subsection{Hong-Ou-Mandel interference}\label{subsec:HOM}
Entanglement swapping is facilitated by a BSM at Charlie, which relies on the interference of indistinguishable photons in the standard optical implementation \cite{braunstein1995measurement}. To evaluate the indistinguishability of the photons from Alice and Bob, we perform HOM interference at Charlie's BS.
We use the same setup as Fig. \ref{fig:setup} except for the removal of the MI's from Alice and Bob, such that each idler mode is measured by a single detector.
Photon pairs are prepared in the state $\ket{\text{TMSV}}_{e}$ by injecting a single RF pulse into each IM at a repetition rate of 200 MHz.
For low mean photon numbers,
 \begin{align}
     \ket{\text{TMSV}}_{e}\approx \sqrt{1-\mu}\ket{0} + \sqrt{\mu}\ket{1,1}_e + \mathcal{O}(\mu^2), \quad \mu \ll 1,\label{eq:hom}
 \end{align}
 neglecting loss. 
 With mean photon numbers per time-bin of $\mu_A = 0.019$ and $\mu_B = 0.015 \ll 1$ for this measurement, each signal mode is approximately in a single photon state conditioned on the detection of its idler mode.
 
 We interfere the signal photons and measure the coincidence events at the output of Charlie's BS. The variable delay line at Bob's input to the BS is used to sweep the relative time-delay, i.e. temporal distinguishability, of Alice and Bob's signal photons  ($\Delta t_{AB}$).  By heralding the signal photons with the detection of the idler photons at Alice and Bob, we measure fourfold coincidence rates for various $\Delta t_{AB}$ over a range of 560 ps. The results are shown in Fig. \ref{fig:hom}a.  Assuming Gaussian temporal profiles of the optical pulses with 1/e temporal duration of $\sigma = 25$ ps, the coincidence rates are fitted proportional to $1-V_{\text{HOM}}\exp(-\Delta t_{AB}^2/2\sigma^2)$, where the HOM visibility is $V_\text{HOM} = (C_d- C_i)/C_d$, with $C_{d(i)}$ denoting the coincidence rates when the photons are made as distinguishable (indistinguishable) as possible. Single photons that are indistinguishable in all degrees of freedom (e.g. temporal, spectral, spatial) would result in a HOM visibility of 100\%. We obtain a HOM visibility of $V_\text{HOM}^{(4)} = 86.7\pm1.8\%$, indicating high indistinguishability of photons from Alice and Bob. The deviation from unity visibility is expected from experimental imperfections including multiphoton contributions and distinguishability in the temporal mode profiles of the photons from Alice and Bob introduced during optical pulse generation.

To glean further information about the quantum interference at Charlie, we also measure threefold and twofold coincidence rates for various $\Delta t_{AB}$.   The HOM visibility depends on the photon statistics of the interfering fields. Without heralding a signal photon by the detection of an idler photon, the state of the signal mode is described by a thermal state,
\begin{align}
\rho_\text{th} = \text{Tr}_{i}\ket{\text{TMSV}}\bra{\text{TMSV}} = \sum_{n=0} \frac{\mu^n}{(1+\mu)^{n+1}}\ket{n}\bra{n} \label{eq:thermal_state}
\end{align}
 where $\text{Tr}_i$ denotes the partial trace over the idler mode of the TMSV.
By heralding only one of the signal photons by the detection of an idler photon at Alice or Bob, we measure threefold coincidence rates corresponding to the interference of a single photon state and thermal state. The twofold coincidence rates at the output of Charlie's BS correspond to the interference of two thermal states. We obtain HOM visibilities of $V_\text{HOM}^{(3B)}=28.4\pm1.2\%$ and $V_\text{HOM}^{(3A)}=22.3 \pm 1.2\%$ for the threefold coincidence rates conditioned on the idler mode at Bob  (Fig. \ref{fig:hom}b) and Alice  (Fig. \ref{fig:hom}c), respectively. The asymmetry in the threefold HOM visibilities is expected due to the difference in the mean photon numbers of Alice and Bob's sources and heralding path efficiencies. For the twofold coincidence rates, we obtain a HOM visibility of $V_\text{HOM}^{(2)}=7.3\%\pm0.2\%$ (Fig. \ref{fig:hom}d). 

We support our measurements with modeling as described in Section \ref{sec:modeling}. 
For the twofold HOM visibility, we obtain an upper bound of 33\% corresponding to the interference of ideal thermal states.
We obtain an upper bound of 50\% for the threefold HOM visibility corresponding to the interference of ideal single photon and thermal states with identical mean photon numbers. Threefold HOM visibilities of up to 100\% could be achieved with unequal mean photon numbers (see Appendix \ref{app:HOM_sweep}).  
Relevant to the entanglement swapping configuration, we find that our fourfold HOM visibility corresponds to a photon indistinguishability of $0.92 \pm 0.02$ (see Appendix \ref{sec:Vhom_app}).
The presence of clear HOM dips and estimation of high photon indistinguishability indicate that our system can perform BSMs suitable for entanglement swapping.

\begin{figure}[htbp!]
  \centering
    \includegraphics[width=\columnwidth]{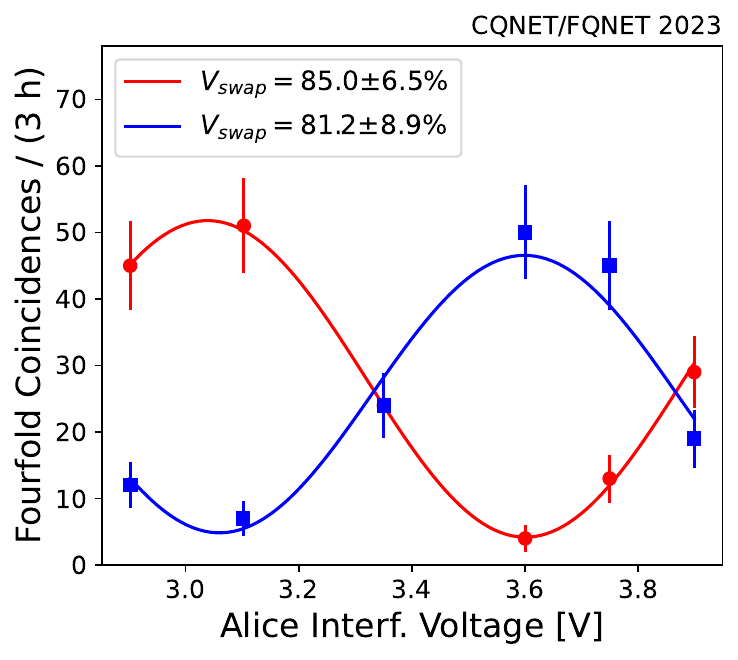}
  \caption{Entanglement swapping of $\ket{\Phi^+}$. The voltage of the Alice's MI is varied to yield a sinusoidal variation of the fourfold coincidence rates for each pairing of output ports of Alice's and Bob's MIs. This yields four sets of fourfold coincidence rates, with two in-phase and two out-of-phase. The in-phase sets are combined (red) and the out-of-phase sets are combined (blue) to obtain two curves. A sinusoidal fit is performed for each combined data set to extract the swapping visibilities of $V_\text{swap}=85.0\pm6.5\%$ (red) and $V_\text{swap}=81.2\pm8.9\%$ (blue). The average of the two visibilities is $\langle V_\text{swap}\rangle= 83.1\pm 5.5\%$.
 }
  \label{fig:swapping}
\end{figure}

\subsection{Entanglement swapping}\label{subsec:teleportation}

After characterization of our system, we perform the entanglement swapping protocol with the setup in Fig. \ref{fig:setup}. We measure the entanglement visibility of idler photons at Alice and Bob conditioned on the BSM at Charlie, resulting in fourfold coincidence rates for each pairing of Alice and Bob's MI output ports. The phase of Alice's MI is varied, and for each interferometric voltage setting the fourfold coincidences are acquired for three hours. 

The results are shown in Fig.  \ref{fig:swapping}.   We obtain two curves by combining fourfold coincidence rates for in-phase pairings of MI outputs (see Fig. \ref{fig:entanglement_visibility}) and observe visibilities of $V_{\text{swap}} = 85.0 \pm 6.5\%$ and $V_{\text{swap}} = 81.2\pm8.9\%$, which surpass the classical bound of $1/3$ required to demonstrate entanglement \cite{peres1996separability}. The average visibility of $\langle V_{\text{swap}}\rangle = 83.1\pm5.5\%$  corresponds to a teleported state fidelity of $\langle F_{\text{swap}}\rangle = 87.3\pm4.1\%$ with respect to $\ket{\Phi^{+}}$ and a violation of the CHSH Bell inequality by $2.25$ standard deviations.
\begin{table}[t]
\caption{\label{tab:qkd}%
Source-independent quantum key distribution error rates. The secret key rate ($R/R_s$) is calculated from the measured error rates in the time ($e_t$) and phase ($e_p$) bases for an error correction efficiency of $\kappa = 1.22$. Error bars on the rates are calculated from the propagation of Poisson statistics.
}
\begin{ruledtabular}
\begin{tabular}{ccc}
\textrm{Time basis ($e_t$)}&
\textrm{Phase basis ($e_p$)}&
\textrm{Secret key rate ($R/R_s$)}\\
\colrule
$0.011 \pm0.011$ & $0.079\pm 0.020$ & $0.50^{+0.18}_{-0.14}$\\

\end{tabular}
\end{ruledtabular}
\end{table}

\subsection{Source-independent quantum key distribution}

Alice and Bob can obtain a secure key by measuring the idler photons in the time basis $\{\ket{e}, \ket{\ell}\}$ and phase basis  $\{(\ket{e}\pm \ket{\ell})/\sqrt{2}\}$ conditioned on the BSM. From the security proof of Koashi and Preskill \cite{koashi2003secure}, the secret key rate for quantum key distribution (QKD) \cite{scherer2011long} with a basis-independent source \cite{ma2007quantum} is,
\begin{align}
    R\geq R_S[1-\kappa H_2(e_t) - H_2(e_p)] \label{eq:secret_key_rate}
\end{align}
where $R_s$ is the sifted key rate, $\kappa$ is the error correction efficiency, $e_t$ is the bit error rate in the time basis, $e_p$ is the bit error rate in the phase basis, and $H_2$ is the binary entropy function, $$H_2(x) = -x\log_2(x)-(1-x)\log_2(1-x).$$ 

With the setup in Fig. \ref{fig:setup}, we measure the error rates for QKD.
For the phase-basis error rate, we measure Alice and Bob's idler photons in the phase basis conditioned on the BSM and accumulate fourfold coincidences for a total of twelve hours. For the time-basis error rate, we remove the MI's to measure Alice and Bob's idler photons in the time basis and accumulate fourfold coincidences for the same period. The results are summarized in Table I. For identical time-basis and phase-basis error rates ($e_t = e_p$), an error rate of less than $11\%$ is required for a non-zero secret key rate. In practice, the phase-basis error rate is higher than the time-basis error rate due to experimental challenges associated with the quantum interference requirements of the phase basis.  We obtain $<10\%$ error rates in both the phase and time bases. The error rate in the phase basis is consistent with the average entanglement swapping visibility of Fig. \ref{fig:swapping}, which estimates $e_p = (1-\langle V_{\text{swap}}\rangle)/2 = 0.085\pm 0.028$.  The secret key rate per sifted key is obtained from Eq. (\ref{eq:secret_key_rate}) with $\kappa = 1.22$. The nonzero secret key rate of $0.50^{+0.18}_{-0.14}$ bits per gate illustrates the suitability of our system for metropolitan-scale quantum key distribution.

\section{Analytical Modeling}\label{sec:modeling}

As discussed in Sec. \ref{sec:results}, experimental implementations of quantum networks introduce nonidealities, such as multiphoton effects, multiple modes, and dark counts, that can degrade the performance of quantum communication protocols in the real world. 

Modeling of quantum networks that can account for all experimental imperfections will elucidate the performance criteria for quantum network components and provide valuable insight for the scale-up of quantum network testbeds towards the quantum internet \cite{khalique2015practical}.
Typically, SPDC-based experiments are modeled in the photon number basis, where the analytical calculations for multimode coincidence probabilities quickly become intractable without low mean photon number approximations. 

In this work, we extend our phase-space-based Gaussian model for quantum teleportation \cite{theory_nikolai} to entanglement swapping \cite{weedbrook2012gaussian}. 
Since the TMSV states produced by SPDC have a Gaussian characteristic function, and all subsequent experimental operations up to detection are described by linear optics, we construct the symplectic matrix that maps the characteristic function of the input state to that of the output state prior to detection \cite{Takeoka2015}. 
By modeling each SNSPD as a threshold detector, whose POVM can be written in terms of Gaussian states, i.e. states with Gaussian characteristic functions, we derive expressions for all detection probabilities in the experiment. 
In this approach, all multiphoton contributions are captured by the characteristic function of the input state, and imperfections such as loss, photon distinguishability, and dark counts can be modeled with symplectic matrices. 
Therefore, we can efficiently compute the output state accounting for all relevant experimental imperfections and obtain exact analytical expressions for the entanglement and HOM visibilities. 
The derivations for HOM and entanglement swapping visibilities as a function of Alice and Bob's mean photon numbers, photon indistinguishability, path efficiencies, imperfect beamsplitter transmittances, and dark count rates are outlined in Appendix \ref{app:theory_calc}. 
Theoretical investigations of the HOM and entanglement swapping visibilities as a function of mean photon numbers and photon indistinguishibilities, with comparison to the data, are presented in Appendix \ref{sec:Vhom_app} and \ref{sec:Vswap_app}.

\begin{figure}[htbp!]
  \centering
    \includegraphics[width=\columnwidth]{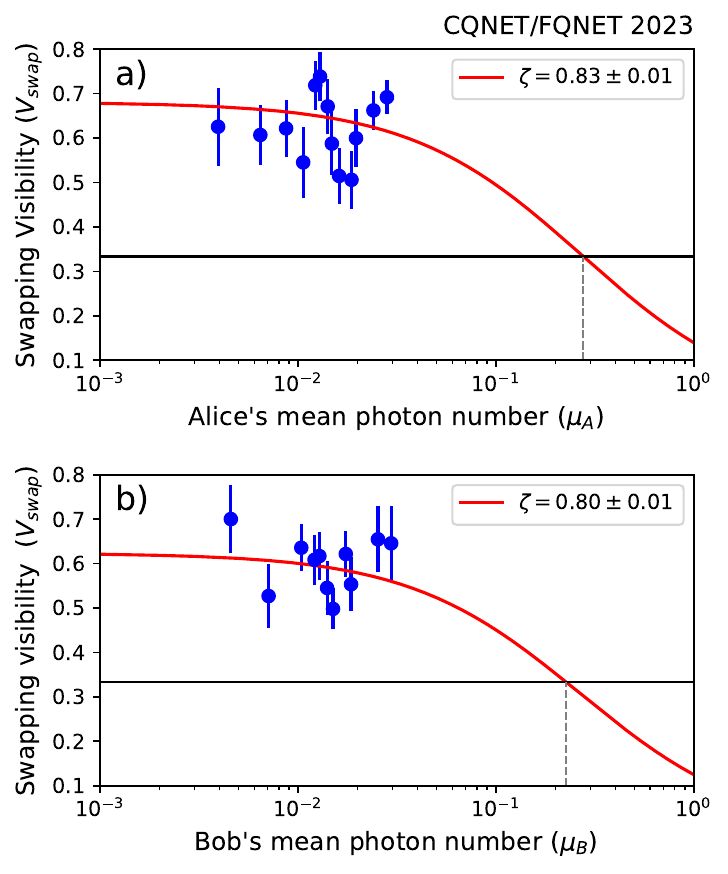}
  \caption{Entanglement swapping visibility as a function of a) Alice's mean photon number ($\mu_A$) and b) Bob's mean photon number ($\mu_B$). 
  The data (blue) are fit to the theoretical model (red) for fixed mean photon numbers of a) $\mu_B = 4.6\times10^{-3}$ and b) $\mu_A = 3.9\times10^{-3}$, with the indistinguishability parameter $\zeta$ as a free parameter. The extracted parameters correspond to indistinguishabilities of a) $\zeta^2 = 0.69\pm0.02$ and b) $\zeta^2 =0.64\pm0.02$. The black lines are the classical bound of $1/3$. 
  } 
  \label{fig:swapping_model_data}
\end{figure}

We experimentally investigate the swapping visibility for asymmetric mean photon numbers of Alice and Bob's sources by fixing Bob's (Alice's) mean photon number 
 and sweeping Alice's (Bob's) mean photon number. The swapping visibility as a function of Alice's and Bob's mean photon numbers are shown in Fig. \ref{fig:swapping_model_data}. The mean photon number is varied by sweeping the gain of the EDFA at Alice or Bob. The EDFAs are remotely controlled by the DAQ subystem to perform automated gain sweeps, enabling long duration data collection and optimization of the system over a range of mean photon numbers. For our repetition rate of 200 MHz, the maximum accessible mean photon number was $0.05$. We fit the data to the swapping model to determine the indistinguishability and obtain $\zeta^2 = 0.69 \pm 0.02$ $(0.64\pm0.02)$ for the sweep over Alice's (Bob's) mean photon number. The swapping visibility exceeds the classical bound up to $\mu_A = 0.28$ and $\mu_B = 0.23$ for Alice's and Bob's sweeps, respectively.

\section{Discussion}\label{sec:discussion}

We demonstrate entanglement swapping between entangled time-bin qubits encoded into 1536.4 nm-wavelength photon pairs with an average fidelity of 87\%, which permits source-independent quantum key distribution with a key rate of 0.5 bits per sifted bit. Our system is interpreted using characteristic function-based analytical modeling that accounts for realistic imperfections such as loss, indistinguishability, and undesired photon-number components. The system is semi-autonomous and uses modular, fiber-coupled, off-the-shelf, and electrically interfaced components, such as modulators and SNSPD systems, that can be straightforwardly replicated for multi-node networks. Nonetheless, our current fidelities and swapping rates of 0.01 Hz are still limited.

Concerning fidelity, our modeling predicts that completely indistinguishable photons will yield a swapping fidelity of 97\% (see Appendix \ref{sec:Vswap_app}). As suggested by the increased indistinguishability with heralding, reducing the multi-mode nature of the pairs by improved control of the photon pair spectra will also reduce $\zeta$. Further improved indistinguishability could be achieved by broader pump bandwidths, e.g., using a mode-locked laser, narrower filtering (at the expense of rates), cavity-enhanced SPDC, or dispersion-engineered sources, in addition to improved setup stability, such as better temperature and polarization control. Even with complete indistinguishability, multi-photon events must be suppressed. Given the mean photon number is quite low at $\sim10^{-3}$, options include replacing the SNSPDs at the BSM with photon-number-resolving (PNR) SNSPDs if allowed by the networking scheme, or using sources of near-deterministic entangled photon pairs based on single emitters, e.g., quantum dots, or multiplexed probabilistic sources, e.g., using SPDC. These approaches could lead to increased effective mean photon number and thereby improve swapping rates, but heralding of multi-photon events must be considered \cite{theory_nikolai, davis2022improved}. Fourfold coincidence detection renders the effect of dark counts negligible.

The current coupling efficiency of $\sim5\%$ per signal or idler channel indeed restricts swapping rates. The loss can be minimized to less than a few dB per signal or idler channel by improved device packaging, using lower-loss components, and splicing. For example, the spectral filters used in this work have a tunable passband which is accompanied by up to 10 dB loss and should be replaced with an alternative, such as a wavelength-division multiplexing (WDM) filter. Alternatively, we could integrate our system onto a chip, for instance using thin-film lithium niobate \cite{zhu2021integrated}. A factor of two in Bell-state measurement efficiency can be gained by projecting onto $\ket{\Psi^+}$ using faster-recovery SNSPDs \cite{valivarthi2014efficient}. The $\sim5\%$ system coupling efficiency is equivalent to $\sim70$ km of single-mode fiber, which suggests our system is already suited for deployment.

An increase in the clock rate will also benefit swapping. Without spatial multiplexing, this can be achieved by accessing the available time-frequency space. A $\sim$100 GHz clock rate is realistic given the demonstration of SNSPDs with ps-level timing resolution and the high bandwidths of electro-optic modulators \cite{prash2021breaking}. This would allow constructing the interferometer on chip, providing superior phase stability, and hence, fidelity. Another approach is to use mode-locked lasers for pumping of the SPDC. This is particularly attractive as the broadband pump will allow extending the multiplexing to the frequency domain, and the use of WDMs can access multiple distinct frequency channels, an approach that co-authors demonstrated previously \cite{mueller2023high}. Frequency multiplexing techniques are compatible with the aforementioned sources that we suggest to increase fidelities and can be extended to repeaters with frequency shifting \cite{sinclair2014spectral}.

With some of the aforementioned improvements, we expect that our system can be deployed for metro-scale networking, demonstrations of quantum hardware interfacing, e.g., with erbium ions, or configured toward sensing protocols, e.g., long-baseline telescopes \cite{gottesman2012telescope}. Our setup is straightforwardly extended to using independent lasers at different locations provided appropriate feedback mechanisms are employed \cite{sun2017entanglement}. Furthermore, our approach can be rendered more cost-effective (e.g., with field-programmable gate arrays replacing the AWG) to realize scaled quantum internet nodes.

\section*{Acknowledgements}\label{sec:ack}
We thank Prof. Anton Zeilenger for suggesting the ``teleportation of entanglement`` term for entanglement swapping in 2023. 

S.D. is partially supported by the Brinson Foundation. R.V., N.L., L.N., C.P., N.S., M.S., S.D, R.Y. and S.X. acknowledge partial support from the Alliance for Quantum Technologies’ (AQT/INQNET) Intelligent Quantum Networks and Technologies research program that was used to build quantum networking infrastructure at Caltech and Fermilab since 2017.  R.V., N.L., L.N., C.P., N.S., M.S.  S.X. and acknowledge partial support from the U.S. Department of Energy, Office of Science, High Energy Physics, QuantISED program grant, under award number DE-SC0019219.  C.P., S.X., A.C., R.V., L.N., M.S., R.Y.,   acknowledge partial support from the U.S. Department of Energy, Office of Science, Advanced Scientific Computing Research, Quantum Communication program grant, under award number (IEQNET) and (AQNET). C.W. received full support from the Office of Science DOE Graduate Student Research Fellowship (SCGSR, 08/2022-08/2023).
 K.T. acknowledges the Samuel P. and Frances Krown SURF fellowship. R.V. and M.S. are also partially supported by the DoE's ASCR QUANT-NET project. The core of this work was conducted between 2022 and 2023 following our publication in PRX Quantum 1, 020317, 2020, with the manuscript prepared in 2024. M.S. is grateful to the Caltech SURF program for partial support of undergraduate research. C.L., K.T  worked on this project alongside their SURF research, and have since moved on to graduate school. K.K. and R.Y. worked as gap-year students and moved on to graduate school as well.



\appendix
\section{Graphical User Interface (GUI)} \label{sec:gui}
\begin{figure}[htbp!]
  \centering
   \includegraphics[width=\columnwidth]{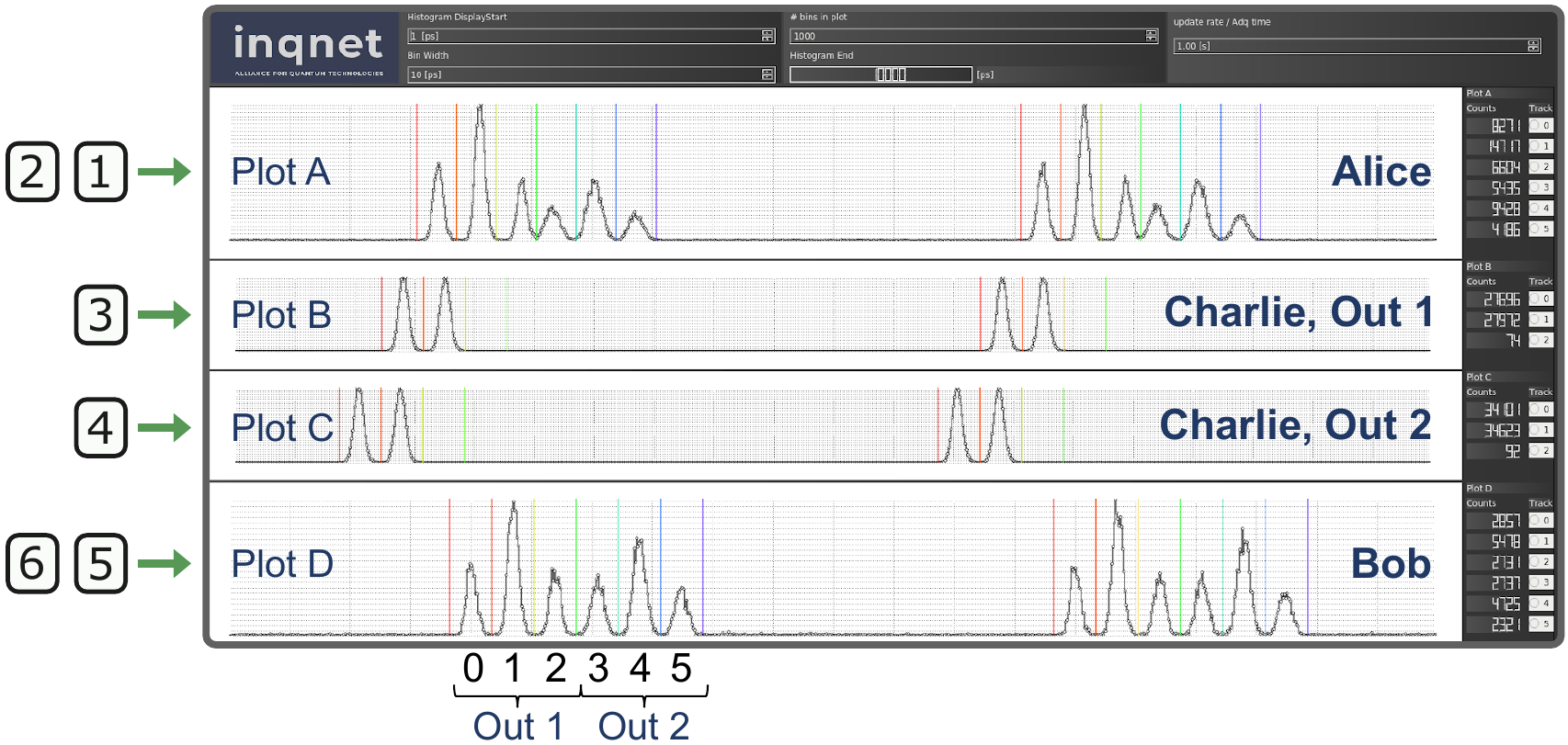}
   \caption{Graphical User Interface (GUI) used to perform real-time data acquisition and analysis. The top (bottom) plot corresponds to the electronically-combined outputs of the detectors at Alice (Bob) and the middle plots correspond to the outputs of each detector at Charlie. Each coincidence window is indicated by a pair of colored bars, which are user-defined and enable tunable temporal filtering.
   }
   \label{fig:daq}
 \end{figure}

We developed a Graphical User Interface (GUI) for the analysis of photon time-of-arrival statistics in quantum networks, see Fig. \ref{fig:daq}. 
The GUI contains four plots corresponding to the time-tags from each channel of the time tagger (TDC). Each plot depicts a histogram of the time-tags relative to the clock signal for a given acquisition time that is set by the user. The histograms update after each acquisition time for live visualization of photon time-of-arrival statistics. 
The GUI supports tunable coincidence windows for up to 10 qubits per clock cycle, enabling reconfigurable coincidence logic. Coincidences can be accumulated over selected coincidence windows for an acquisition time set by the user. All detection events are recorded to a MySQL database after each acquisition time, allowing for automated data collection, real-time monitoring, and big data storage accessible throughout the network over long-term experimental operation.

\begin{table}[htbp!]
\begin{ruledtabular}
\begin{tabular}{ccccccc}
&
\textrm{$\mu_A$}&
\textrm{$\mu_B$}&
\textrm{$\eta_{Ai}$}&
\textrm{$\eta_{As}$}&
\textrm{$\eta_{Bs}$}&
\textrm{$\eta_{Bi}$}\\
\colrule
(a)&$0.019$ & $0.015$ & $0.067$ & $0.10$&$0.11$&$0.072$\\
(b)&$0.0047$ & $0.0042$ & $0.017$ & $0.048$&$0.066$&$0.020$\\
(c)&--  & $0.0046$ & $0.026$ & $0.072$&$0.076$&$0.022$\\
(d)&$0.0039$  & -- & $0.031$ & $0.078$&$0.076$&$0.022$
\end{tabular}
\end{ruledtabular}
\caption{\label{tab:exp_params}%
Experimental parameters for the Hong-Ou-Mandel (HOM) and swapping configurations. $\mu_{A(B)}$ is the mean photon number of Alice's (Bob's) photon pair source, $\eta_{Ai}$ is Alice's idler path efficiency, $\eta_{As}$ is Alice's signal path efficiency, $\eta_{Bs}$ is Bob's signal path efficiency, and $\eta_{Bi}$ is Bob's idler path efficiency. (a) HOM interference measurements in Sec. \ref{subsec:HOM}, (b) entanglement swapping measurements in Sec. \ref{subsec:teleportation}, and entanglement swapping measurements used to test the model in Sec. \ref{sec:modeling} with (c) $\mu_A$ varied while $\mu_B$ is fixed and (d) $\mu_B$ varied while $\mu_A$ fixed.
}
\end{table}

\section{Experimental characterization}
The mean photon numbers of Alice and Bob's photon pair sources and the path efficiencies of Alice and Bob's signal and idler paths for various experimental configurations are reported in Table \ref{tab:exp_params}. These parameters are substituted into the analytical expressions for the HOM and swapping visibilities that are derived in Appendix \ref{app:theory_calc} to generate the theoretical models in Sec. \ref{sec:modeling} and estimate the photon indistinguishabilities for the HOM and swapping experiments.
The mean photon numbers are determined from the coincidence-to-accidental ratio, and the path efficiencies (see Fig. \ref{fig:setup}) are determined from the ratio of coincidence to single-photon detection rates \cite{klyshko1980use}. 
The path efficiencies include the effects of component insertion loss, detection efficiency, and additional loss due to frequency entanglement and heralding \cite{theory_nikolai}.
For the entanglement swapping measurements (b-d), the idler path efficiencies are obtained by summing the efficiencies calculated for each output of the interferometers.

\begin{figure*}[htbp!]
  \centering
    \includegraphics[width=0.9\textwidth]{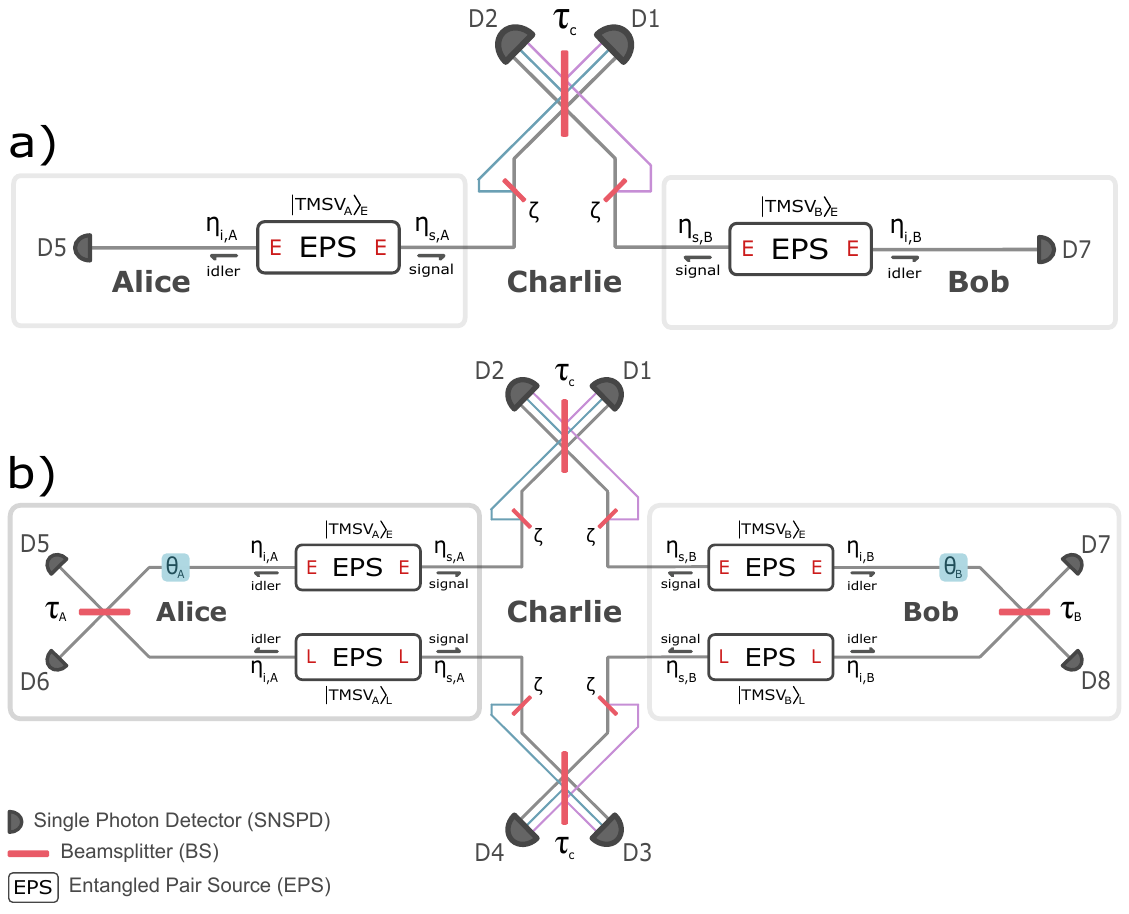}
  \caption{Theoretical model setups for a) Hong-Ou-Mandel interference and b) entanglement swapping. In the setups, $E$ and $L$ denote the early and late modes, respectively. $\eta_{Ai}$ is Alice's idler path efficiency, $\eta_{As}$ is Alice's signal path efficiency, $\eta_{Bs}$ is Bob's signal path efficiency, and $\eta_{Bi}$ is Bob's idler path efficiency. $\tau_{A(B)}$ is a transmittance accounting for imperfect interferometric visibility of Alice's (Bob's) MI, with $\tau_{A(B)}=1/\sqrt{2}$ corresponding to ideal interferometric interference. $\theta_{A(B)}$ is the phase setting of Alice's (Bob's) MI. $\tau_C$ is the transmittance of the beamsplitter at Charlie and $\zeta$ is the photon indistinguishability parameter, where $\zeta^2$ represents the fraction of modal overlap of the photons interfering at Charlie's beamsplitter.}
  \label{fig:model_setup}
\end{figure*}
 \section{Characteristic function approach}\label{app:theory_calc}

The models for the HOM and entanglement swapping experiments are summarized in Fig. \ref{fig:model_setup}a and b, respectively.
We follow the approach proposed in Ref. \cite{Takeoka2015}, which has been applied to quantum teleportation \cite{valivarthi2020teleportation, theory_nikolai} and heralded single photon source experiments \cite{davis2022improved}. 
Using the notation of Ref. \cite{davis2022improved}, the characteristic function for a Gaussian state of an $N$-mode bosonic system is
\begin{align}
    \chi(\xi) = \exp(-\frac{1}{4}\xi^T \gamma \xi - i d^T\xi), \label{eq: char_fxn}
\end{align}
where $\xi \in R^{2N}$, $d$ is the displacement vector, and $\gamma$ is the covariance matrix. States that can be described by Eq. \ref{eq: char_fxn}, including the vacuum, coherent, thermal, single- and two-mode squeezed states, are fully characterized by their displacement vector and covariance matrix. For a TMSV, the displacement vector is the null vector and the covariance matrix is given by
\begin{align}
    \gamma_{\text{TMSV}}(\mu) &= \begin{bmatrix}
        \mathbf{A} & \mathbf{B}\\
        \mathbf{B} & \mathbf{A}
    \end{bmatrix},\\\nonumber
    \mathbf{A} &= \begin{bmatrix}
        1+2\mu & 0\\
        0 & 1+2\mu
    \end{bmatrix},\\\nonumber
    \mathbf{B} &= \begin{bmatrix}
        2\sqrt{\mu(\mu+1)} & 0\\
        0 & -2\sqrt{\mu(\mu+1)} 
    \end{bmatrix},
\end{align}
in block matrix form, where $\mu$ is the mean photon number. 
For the HOM measurements, the input state is modeled as a tensor product of TMSV states from Alice and Bob's sources in the early (e) temporal modes,
\begin{align}
    \ket{\Psi_{in}} = \ket{\text{TMSV}}_{A,e}\ket{\text{TMSV}}_{B,e}, \nonumber
\end{align}
which has the characteristic function,
\begin{align}
    &\chi_{in} (\xi) = \exp(-\frac{1}{4}\xi^T \gamma_{in}(\mu_{A},\mu_{B}) \xi),\label{eq:chi_in}\\
    &\gamma_{in}(\mu_A, \mu_B) = \gamma_{\text{TMSV}}(\mu_{A,e})  \oplus \gamma_{\text{TMSV}}(\mu_{B,e}),\nonumber
\end{align}
where $\gamma_{in}(\mu_A, \mu_B)$ is the input covariance matrix and $\mu_{A} = \mu_{A,e}$,  $\mu_{B} = \mu_{B,e} $ are the mean photon numbers for Alice and Bob's pair sources, respectively. 
For the entanglement swapping experiment, the input is modeled as a tensor product of TMSV states from Alice and Bob's sources in the early ($e$) and late ($\ell$) temporal modes, 
\begin{align}
    \ket{\Psi_{in}} = \ket{\text{TMSV}}_{A,e}\ket{\text{TMSV}}_{A,\ell}
\ket{\text{TMSV}}_{B,e}\ket{\text{TMSV}}_{B,\ell}, \nonumber
\end{align}
which has the characteristic function of Eq. \ref{eq:chi_in} with input covariance matrix
\begin{align}
    \gamma_{in}(\mu_A, \mu_B) =& \gamma_{\text{TMSV}}(\mu_{A,e}) \oplus \gamma_{\text{TMSV}}(\mu_{A,\ell}) \\&\oplus \gamma_{\text{TMSV}}(\mu_{B,e}) \oplus \gamma_{\text{TMSV}}(\mu_{B,\ell}), \nonumber
\end{align}
where we take $\mu_A = \mu_{A,e}= \mu_{A,\ell}$ and $\mu_B = \mu_{B,e}= \mu_{B,\ell}$. We note that spectral impurities can be accounted for by modeling the input state with TMSV states in multiple Schmidt modes \cite{davis2022improved}. 

As mentioned in Sec. \ref{sec:modeling}, the operations of the experiment up to detection can be modeled with linear optical transformations on the spatiotemporal input modes. Linear optics preserve the form of Gaussian characteristic functions, i.e. they map a Gaussian state onto another Gaussian state, by a symplectic transformation of the displacement vector and covariance matrix,
\begin{align}
    d^\prime &= S^T d\\
    \gamma^\prime &= S^T \gamma S,
\end{align}
where $S$ is a symplectic matrix. In the experiment, all transformations on the input state, such as Charlie's BS and the interferometers, can be constructed from beamsplitter and phase shifter transformations. The symplectic matrix $S_{\text{BS}}$ for a beamsplitter is,  
\begin{align}
    S_{\text{\text{BS}}}(t) &= \begin{bmatrix}
        \mathbf{T} & \mathbf{R}\\
        \mathbf{R} & \mathbf{T}
    \end{bmatrix},\\\nonumber
    \mathbf{T} &= \begin{bmatrix}
        t & 0\\
        0 & t
    \end{bmatrix},\\\nonumber
    \mathbf{R} &= \begin{bmatrix}
        0 & -\sqrt{1-t^2}\\
        \sqrt{1-t^2} & 0 
    \end{bmatrix},
\end{align}
and the symplectic matrix $S_\text{P}$ for a phase shifter is the rotation matrix, 
\begin{align}
    S_\text{P}(\theta) = 
    \begin{bmatrix}
        \cos{\theta}& -\sin{\theta}\\
        \sin{\theta} & \cos{\theta}
    \end{bmatrix}.
\end{align}

Each interferometer is modeled as a phase shifter followed by a beamsplitter that interferes the early and late idler modes of an EPS. Optical loss is implemented by mixing an input mode with a virtual vacuum mode through a beamsplitter transformation with transmittance $\sqrt{\eta}$, where $\eta$ is the transmission efficiency. Photon indistinguishability is also modeled with a beamsplitter transformation as detailed in Ref. \cite{valivarthi2020teleportation}, where the indistinguishability parameter $\zeta$ is the transmittance of a virtual beamsplitter that mediates the modal interference of input fields to Charlie's BS. The indistinguishability $\zeta^2$ characterizes the amount of modal overlap of incoming photons, with $\zeta^2 = 1$ for photons that are indistinguishable and $\zeta^2 = 0$ for photons that are distinguishable in all degrees of freedom.

After constructing the overall symplectic matrix for the experiment $S_{\text{exp}}$, the coincidence probabilities are found in terms of the output covariance matrix $\gamma_{\text{out}} = S_{\text{exp}}^T\gamma_{\text{in}} S_{\text{exp}}$ for the output state $\hat{\rho}_\text{out}$.
Given the output state, the probability for a measurement outcome with a measurement operator $\hat{\Pi}$ is, 
\begin{align}
\operatorname{Tr}[\hat{\rho}_\text{out} \hat{\Pi}]=\left(\frac{1}{2 \pi}\right)^N \int d x^{2 N} \chi_\text{out}(x) \chi_{\Pi}(-x), \label{eq:gaussian_trace}
\end{align}
where $\chi_{\Pi}(x)$ is the characteristic function of the measurement operator.
For a threshold detector, the measurement operators are, 
\begin{align}
    \hat{\Pi}_{\text{no event}} &=\ket{0}\bra{0}, \label{eq:Pi_no_event}\\
    \hat{\Pi}_{\text{event}} &=\hat{I}-\hat{\Pi}_{\text{no event}} \label{eq:Pi_event},
\end{align}
where $\hat{I}$ is the identity matrix. Dark counts can be accounted for by taking $\ket{0}\bra{0}\rightarrow (1-\nu)\ket{0}\bra{0}$, where $\nu$ is the dark count probability of a detector. All coincidence probabilities in the experiments can be calculated from \eqref{eq:gaussian_trace}-\eqref{eq:Pi_event}. For example, the fourfold coincidence probabilities for the fourfold HOM visibility (see Fig. \ref{fig:model_setup}a) are calculated as
\begin{align}
    &P_{5217} = \text{Tr}[\hat{\rho}_{\text{out}}\hat{\Pi}_{\text{HOM}}^{(4)}], 
\end{align}
where the measurement operator $\hat{\Pi}_{\text{HOM}}^{(4)}$ is,
\begin{align}
    &\hat{\Pi}_{\text{HOM}}^{(4)}=(\hat{I}^{(2)} -\ket{0}\bra{0}_{\text{D5}})\otimes(\hat{I}^{(2)} -\ket{0}\bra{0}_{\text{D2}})\nonumber\\
    &\otimes (\hat{I}^{(2)} -\ket{0}\bra{0}_{\text{D1}})\otimes (\hat{I}^{(2)} -\ket{0}\bra{0}_{\text{D7}}).
\end{align}
Since the vacuum state has a Gaussian characteristic function, \eqref{eq:gaussian_trace} reduces to linear combinations of Gaussian integrals that simplify using,
\begin{align}
    Tr[\hat{\rho}_{\text{out}}\ket{0}\bra{0}^{\otimes N}] = \frac{2^N}{\sqrt{\text{det}(\hat{I}^{(N)}+\gamma_{\text{out}}^{(N)})}}, \label{eq:vac_trace}
\end{align}
where $N$ is the number of modes being measured, $\ket{0}\bra{0}^{\otimes N} = \ket{0}\bra{0}\otimes \cdots \otimes \ket{0}\bra{0}$ denotes the tensor product of the vacuum operator over the $N$ modes, $\hat{I}^{(N)}$ is the $N$ by $N$ identity matrix, and $\gamma_{\text{out}}^{(N)}$ is the reduced output covariance matrix obtained by tracing all modes but those that are measured.

The coincidence probabilities used to compute the HOM and entanglement swapping visibilities are found from \eqref{eq:gaussian_trace} and \eqref{eq:vac_trace} in terms of determinants of the covariance matrix of output state prior to detection. 
We obtain a 16 by 16 and 32 by 32 covariance matrix for the output states of the HOM and entanglement swapping models, respectively, yielding analytical expressions with a large number of terms. For simplification, we set the dark count rates, which had a negligible effect in the experiments, to zero and focus on the impact of multiphoton events and photon indistinguishability. 
We also set ideal transmittances of $1/\sqrt{2}$ for the beamsplitters at Charlie and inside the interferometers. The impact of imperfect beamsplitter transmittances on entanglement visibility is analyzed in the Supplementary Material of Ref. \cite{Mueller:2023uoj} in this formalism. 

\subsubsection{Hong-Ou-Mandel interference model}
For the HOM model, the twofold ($V_{\text{HOM}}^{(2)}$), threefold ($V_{\text{HOM}}^{(3A)}$, $V_{\text{HOM}}^{(3B)}$), and fourfold ($V_{\text{HOM}}^{(4)}$) HOM visibilities are calculated as,
\begin{align}
    V_{\text{HOM}}^{(2)} &= \frac{P_{21}(\zeta = 0)-P_{21}(\zeta = \zeta_{\text{max}})}{P_{21}(\zeta = 0)}\label{eq:VHOM2},\\
    V_{\text{HOM}}^{(3A)} &= \frac{P_{521}(\zeta = 0)-P_{521}(\zeta = \zeta_{\text{max}})}{P_{521}(\zeta = 0)}\label{eq:VHOM3A},\\
    V_{\text{HOM}}^{(3B)} &= \frac{P_{217}(\zeta = 0)-P_{217}(\zeta = \zeta_{\text{max}})}{P_{217}(\zeta = 0)}\label{eq:VHOM3B},\\
    V_{\text{HOM}}^{(4)} &= \frac{P_{5217}(\zeta = 0)-P_{5217}(\zeta = \zeta_{\text{max}})}{P_{5217}(\zeta = 0)},\label{eq:VHOM4}
\end{align}
where $P_{21}(\zeta)$ is the twofold coincidence probability, $P_{521}(\zeta)$ and $P_{217}(\zeta)$ are the threefold coincidence probabilities, and $P_{5217}(\zeta)$ is the fourfold coincidence probability.
In \eqref{eq:VHOM2}-\eqref{eq:VHOM4}, $\zeta=0$ and $\zeta = \zeta_{\text{max}}$ correspond to maximum photon distinguishability and indistinguishability, respectively.
 For identical mean photon numbers $\mu = \mu_A = \mu_B$ and identical path efficiencies $\eta$,  the analytical expressions for the HOM visibilities are,
\begin{align}
    V_{HOM}^{(2)}=\frac{8\zeta_{\text{max}}^2(1+ \eta \mu)^2}{\left(6+6 \eta \mu+\eta^2 \mu^2\right)\left(4+4 \eta \mu+\left(1-\zeta_{\text{max}}^2\right) \eta^2 \mu^2\right)}, \label{eq:v2hom_app}
\end{align}
\begin{widetext}
\begin{align}
    &V_{HOM}^{(3A)}=1-\frac{\frac{1+\eta\mu+\eta^2\mu^2}{(1+\eta\mu)^2}+\frac{1}{( 1+\eta \mu)\left(-1-2 \eta \mu+\eta^2 \mu\right)}+\frac{8}{-4-4 \eta \mu+\left(-1+\zeta_\text{max}^2\right) \eta^2 \mu^2}+\frac{8}{2(1+\eta \mu)\langle 2+\eta \mu)+(-1+\eta) \eta \mu\left(-2+\left(-1+\zeta_\text{max}^2\right) \eta \mu\right)}}{\frac{1+\eta\mu+\eta^2\mu}{(1+\eta\mu)^2}-\frac{8}{(2+\eta \mu)^2}+\frac{8}{(2+\eta \mu)\left(2+3 \eta \mu-\eta^2 \mu\right)}+\frac{1}{(1+\eta \mu)\left(-1-2 \eta \mu+\eta^2 \mu\right)}},\label{eq:v3hom_app}\\
    &V_{HOM}^{(3B)} = V_{HOM}^{(3A)},\\
    &V_{HOM}^{(4)} = 1-\Bigg[1+\frac{2}{(1+\eta\mu)^2}-\frac{2}{1+\eta\mu}+\frac{-1+\eta(-3+2\eta)\mu}{(1+\eta\mu)(-1+(-2+\eta)\eta\mu)^2}+\frac{8}{-4+\eta\mu(-4+(1+\zeta_\text{max}^2)\eta\mu)}+\label{eq:v4hom_app}\\\nonumber
    &\frac{16}{4+\eta\mu(8-2\eta+(3+\zeta_\text{max}^2(-1+\eta)-\eta)\eta\mu)}+\frac{8}{(2+\eta\mu(3-\zeta_\text{max})+\eta^2\mu(\zeta_\text{max}-1))(-2-\eta\mu(3+\zeta_\text{max})+\eta^2\mu(\zeta_\text{max}+1))}\Bigg]\Bigg/\\\nonumber
    &\Bigg[\frac{1+\eta^2\mu^2}{(1+\eta\mu)^2}-\frac{8}{(2+\eta\mu)^2}-\frac{8}{(-2+(-3+\eta)\eta\mu)^2} - \frac{16}{(2+\eta\mu)(-2+(-3+\eta)\eta\mu)}+\frac{1}{(-1+(-2+\eta)\eta\mu)^2}+\\&\frac{2}{(1+\eta\mu)(-1+(-2+\eta)\eta\mu)}\Bigg].\nonumber
\end{align}
\end{widetext}

\subsubsection{Entanglement swapping model}
For the entanglement swapping model, the entanglement swapping visibility is calculated as,
\begin{align}
    V_\text{SWAP}  = \frac{P_{1467}(\theta_A = 0, \theta_B = 0)-P_{1467}(\theta_A = \pi, \theta_B = 0)}{P_{1467}(\theta_A = 0, \theta_B = 0)+P_{1467}(\theta_A = \pi, \theta_B = 0)},
\end{align}
where $P_{1467}(\theta_A, \theta_B )$ is the fourfold coincidence probability, corresponding to the coincidence rate of Alice and Bob conditioned on the BSM, and $\theta_{A(B)}$ is the phase setting for the interferometer at Alice (Bob). For identical mean photon numbers $\mu = \mu_A = \mu_B$, unit path efficiencies, and perfect indistinguishability $\zeta = 1$, 
the analytical expression for the entanglement swapping visibility is,
\begin{widetext}
    \begin{align}
        V_\text{SWAP}&=\frac{-\frac{4}{4+12\mu+13\mu^2+6\mu^3+\mu^4}+\frac{16}{16+48\mu + 48\mu^2+16\mu^3}}{2+\frac{4}{(1+\mu)^2}-\frac{8}{1+\mu}-\frac{16}{2+5\mu+4\mu^2+\mu^3}+\frac{4}{4+12\mu+13\mu^2+6\mu^3+\mu^4}+\frac{32}{\sqrt{16+56\mu+73\mu^2+42\mu^3+9\mu^4}}+\frac{16}{16+48\mu + 48\mu^2+16\mu^3}}.
    \end{align}
\end{widetext}

\section{HOM interference visibility} \label{sec:Vhom_app}

\begin{figure}[htbp!]
  \centering
    \includegraphics[width=\columnwidth]{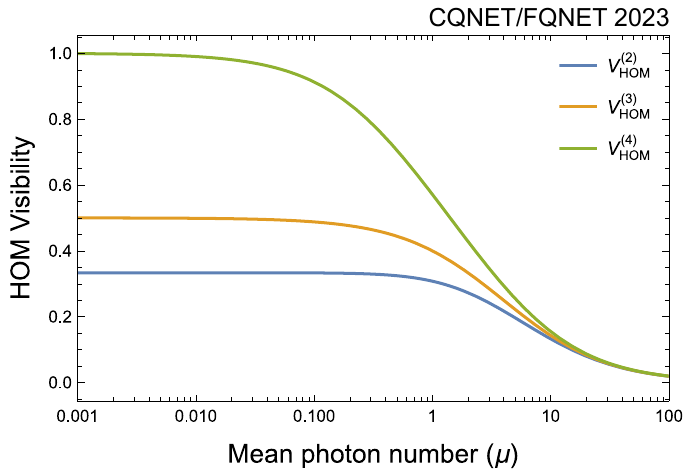}
  \caption{Hong-Ou-Mandel (HOM) visibilities as a function of mean photon number. The solid lines are the theoretical models for the fourfold HOM visibility (green), threefold HOM visibilities (yellow), and twofold HOM visibility (blue) with identical mean photon numbers ($\mu = \mu_A = \mu_B$), unit path efficiencies, and unity indistinguishability. 
  }
  \label{fig:hom_vs_mu}
\end{figure}

\begin{figure}[htbp!]
  \centering
    \includegraphics[width=\columnwidth]{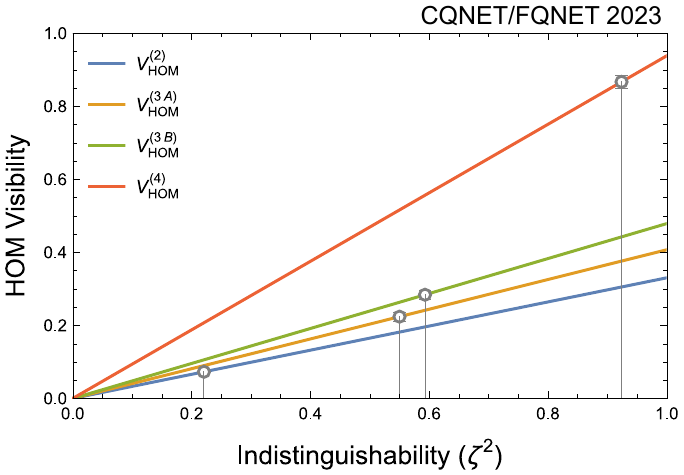}
  \caption{Hong-Ou-Mandel (HOM) visibilities as a function of indistinguishability. The solid lines are the models for the fourfold HOM visibility (red), threefold HOM visibility conditioned on Bob (green), threefold HOM visibility conditioned on Alice (green), and twofold HOM visibility (blue) for the experimentally characterized path efficiencies and mean photon numbers. The data are indicated with circular markers. }
  \label{fig:hom_vs_zeta}
\end{figure}

The HOM visibilities as a function of mean photon number ($\mu = \mu_A = \mu_B$) and 
photon indistinguishability ($\zeta^2$) are shown in Fig. \ref{fig:hom_vs_mu} and 
is shown in Fig. \ref{fig:hom_vs_zeta}, respectively. 
From comparison of the experimental HOM visibilities of Fig. \ref{fig:hom} with the models in Fig. \ref{fig:hom_vs_zeta}, we find indistinguishabilities of  $\zeta^2 = 0.22\pm 0.01$ for the twofold HOM visibility, $\zeta^2 = 0.55\pm 0.03$ for the threefold HOM visibility conditioned on Alice's idler photon, $\zeta^2 = 0.59\pm 0.03$ for the threefold HOM visibility conditioned on Bob's idler photon, and $\zeta^2 = 0.92 \pm 0.02$ for the fourfold HOM visibility measurements. 
Notice that the indistinguishability increases with the number of coincidentally detected photons.
As discussed in previous analyses of quantum teleportation \cite{theory_nikolai}, this is due to frequency entanglement of the photon pairs and spectral filtering.
Filtering and detection of the idler photon reduces the number of spectral modes that are correlated with the signal, hence fourfold detection better approximates single mode behavior.

\subsection{Taylor expansion}

The HOM visibility expressions to lowest order of the multivariate Taylor expansion in $\mu_A$, $\mu_B$ are, 
\begin{align}
V^{(2)}_{\text{HOM}} &\approx \frac{\eta_{s,B}\mu_B/\eta_{s,A}\mu_A}{1+\mu_B\eta_{s,B}/\mu_A\eta_{s,A}+(\mu_B\eta_{s,B}/\mu_A\eta_{s,A})^2}\zeta^2 \label{eq:V2HOM_taylor}\\
V^{(3A)}_{\text{HOM}} &\approx \frac{(\eta_{s,B}\mu_B/\eta_{i,A}\mu_A)}{(2-\eta_{s,A}) + (\eta_{s,B}\mu_B/\eta_{i,A}\mu_A)}\zeta^2\label{eq:V3AHOM_taylor}\\
V^{(3B)}_{\text{HOM}} &\approx \frac{1}{1+(2-\eta_{s,B})(\eta_{i,B}\mu_B/\eta_{s,A}\mu_A)}\zeta^2\label{eq:V3BHOM_taylor}\\
V^{(4)}_{\text{HOM}} &\approx \zeta^2\label{eq:V4HOM_taylor}
\end{align}
Consider the upper bounds of Eq. \ref{eq:V2HOM_taylor}-\ref{eq:V4HOM_taylor} for ideal photon indistinguishability, $\zeta^2 = 1$.
The maximum twofold HOM visibility is $1/3$ at $\eta_{s,B}\mu_B = \eta_{s,A}\mu_A$, as expected for the interference of two thermal states. 
 The threefold visibility $V^{(3A)}_{\text{HOM}}$ heralded on Alice approaches unity for $\eta_{s,B}\mu_B \gg \eta_{i,A}\mu_A$, and the threefold visibility $V^{(3B)}_{\text{HOM}}$ heralded on Bob approaches unity for $\eta_{s,A}\mu_A \gg \eta_{i,B}\mu_B$. The maximum fourfold visibility is unity. 
\subsection{Optimization of HOM visibility} 
\begin{figure*}[htbp!]
        \centering
        \begin{subfigure}[b]{0.475\textwidth}
            \centering
            \includegraphics[width=\textwidth]{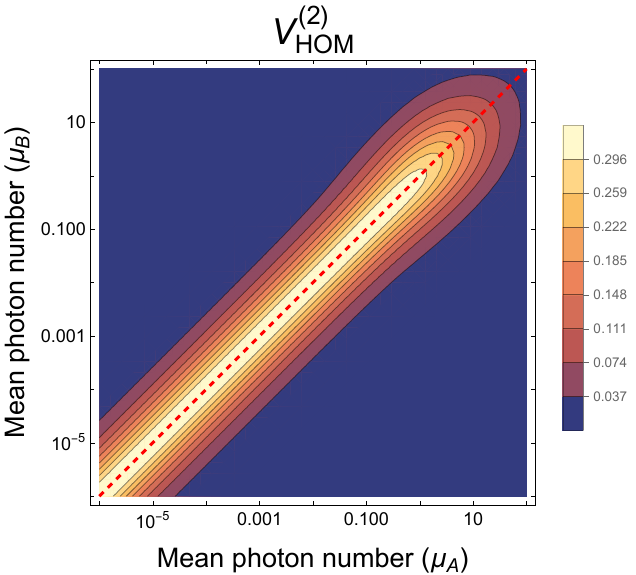}
            \caption
            {{\small Twofold HOM interference visibility of Alice and Bob's signal modes, corresponding to the interference of two thermal states.\\}}    
            \label{fig:V2HOM_sweep}
        \end{subfigure}
        \hfill
        \begin{subfigure}[b]{0.475\textwidth}  
            \centering 
            \includegraphics[width=\textwidth]{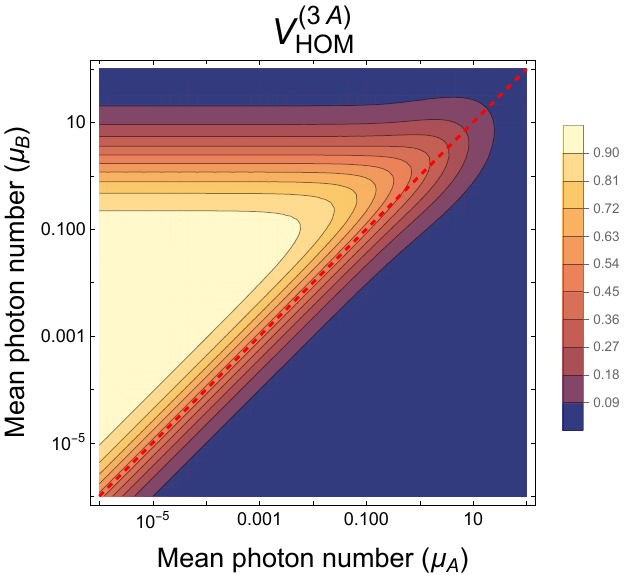}
            \caption[]%
            {{\small Threefold HOM interference visibility of Alice and Bob's signal modes conditioned on Alice's idler mode, corresponding to the interference of a heralded single photon state and thermal state.}}    
            \label{fig:mean and std of net24}
        \end{subfigure}
        \vskip\baselineskip
        \begin{subfigure}[b]{0.475\textwidth}   
            \centering 
            \includegraphics[width=\textwidth]{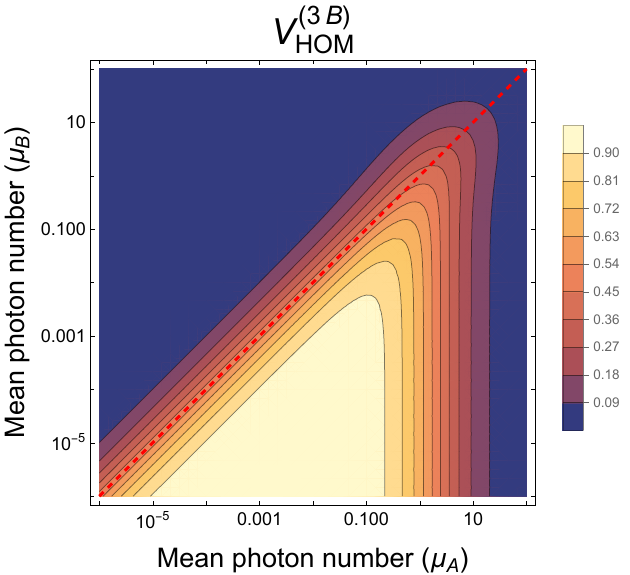}
            \caption[]%
            {{\small Threefold HOM interference visibility of Alice and Bob's signal modes conditioned on Bob's idler mode, corresponding to the interference of a thermal state and heralded single photon  state.}}    
            \label{fig:mean and std of net34}
        \end{subfigure}
        \hfill
        \begin{subfigure}[b]{0.475\textwidth}   
            \centering 
            \includegraphics[width=\textwidth]{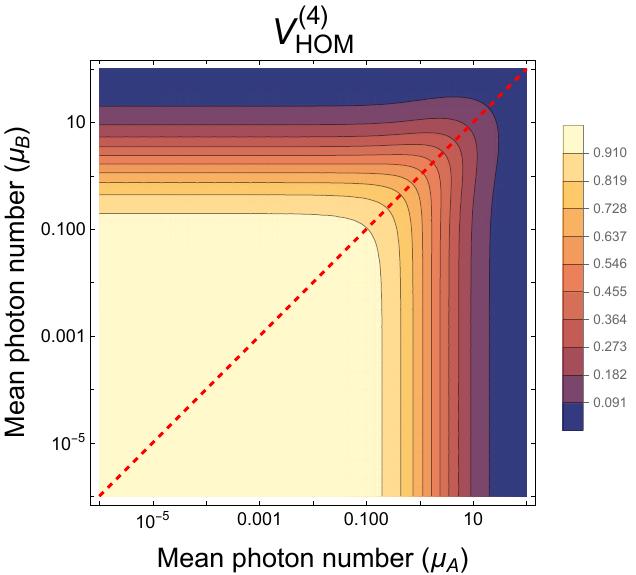}
            \caption[]%
            {{\small Fourfold HOM interference visibility of Alice and Bob's signal modes conditioned on Alice and Bob's idler modes, corresponding to the interference of two heralded single photon states.}}    
            \label{fig:mean and std of net44}
        \end{subfigure}
        \caption[ The average and standard deviation of critical parameters ]
        {\small HOM interference visibilities plotted as a function of Alice and Bob's mean photon numbers for ideal path efficiencies ($\eta_i, \eta_s = 1$) and photon indistinguishability ($\zeta = 1$). The red dashed lines correspond to $\mu_A = \mu_B$, and we include in all plots to facilitate comparison to the two-fold HOM visibility plot.} 
        \label{fig:HOM_sweeps}
    \end{figure*}

The complete HOM visibility expressions, without approximation, are plotted for identical $\mu = \mu_A = \mu_B$ in Fig. \ref{fig:hom_vs_mu} and various $\mu_A$ and $\mu_B$ in Fig. \ref{fig:HOM_sweeps} for unit photon indistinguishability and path efficiencies.  
The behaviors of the plots extend those produced in a previous analysis of quantum teleportation \cite{theory_nikolai}.

For the case of two-fold HOM interference, the visibility is maximized for the case in which mean photon numbers of the two input thermal fields match.
For low mean photon number, the maximum visibility is $1/3$, which corresponds to $\mu_{A}=\mu_{B}$, and is the global maximum.
This is consistent with the linear and symmetric ridge-like topography in Fig. \ref{fig:HOM_sweeps}a.
The maximum value is not unity due to $n=2$ photon states and vacuum input into the beamsplitter for $\mu_A,\mu_B<<1$.
For mean photon numbers approaching one and beyond, the maximum visibility is reduced and the range of mean photon numbers to maximize the visibility increases due to interference from of higher photon number terms.

Due to heralding, the three-fold HOM visibility plots in Fig. \ref{fig:HOM_sweeps}b and c have a plateau-like topography which extend the range of mean photon numbers that allow reaching maximum interference visibility.
The theoretical maximum visibility is unity also due to heralding.
In the case of conditional detection of photons in Alice idler mode (Fig. \ref{fig:HOM_sweeps}b), the threshold at $\mu_{B}\sim1$ is due to $n=2$ events from $\mu_{B}$ interfering with heralded single photons and reducing the maximum visibility.
Provided $\mu_{B},\mu_{A}<<1$, the visibility is maximized independent of the probability of generating a photon in Bob's signal mode because a single photon is always in Alice's signal mode and three-fold detection is performed.
In the case $\mu_{A}$ is increased and starts to approach $\mu_{B}$, the relative probability of heralding a multi-photon term in Alice's signal mode increases, which decreases the visibility, and leads to the threshold topography along the diagonal.
The visibility is not maximized for $\mu_{A}=\mu_{B}$, in this case reaching up to 1/2 (see Fig. \ref{fig:hom_vs_mu}), because heralding increases the effective mean photon number of Alice's signal mode.
In this case, a lower value of $\mu_{A}$ is required to reach maximum visibility compared to two-fold HOM interference, effectively shifting the ridge to the left in Fig. \ref{fig:HOM_sweeps}b compared to that in Fig. \ref{fig:HOM_sweeps}a.
Note that the gradient is smaller at $\mu_{B}\sim1$ due to the presence of a single photon in Alice's signal mode. 
This renders the contributions of higher order terms to be less detrimental to the visibility than along the diagonal where effective mean photon numbers are balanced, and higher order terms contribute in both Alice and Bob's signal mode.
The same arguments apply to explain the topography in Fig. \ref{fig:HOM_sweeps}c in which Bob's idler is detected.
Overall, the difference in number distributions explains the symmetry breaking around the diagonal.

In the case of four-fold detection, the range of mean photon numbers that yield maximum visibility is extended further since both Alice and Bob herald single photons at the beamsplitter.
Since both Alice and Bob detect their idlers, the plot in Fig. \ref{fig:HOM_sweeps}d  is symmetric about the diagonal and has a topography akin to combining both three-fold plots together. 
The small-gradient thresholds remain because the effective number distributions that are heralded are the same as the three-fold case.
In the case of high mean photon numbers, matching of the effective mean photon number is required to maximize until higher number terms dominate at beyond $\mu_{A}=\mu_{B}=1$.

\label{app:HOM_sweep}

    \begin{figure*}[htbp!]
  \centering
    \includegraphics[width=\textwidth]{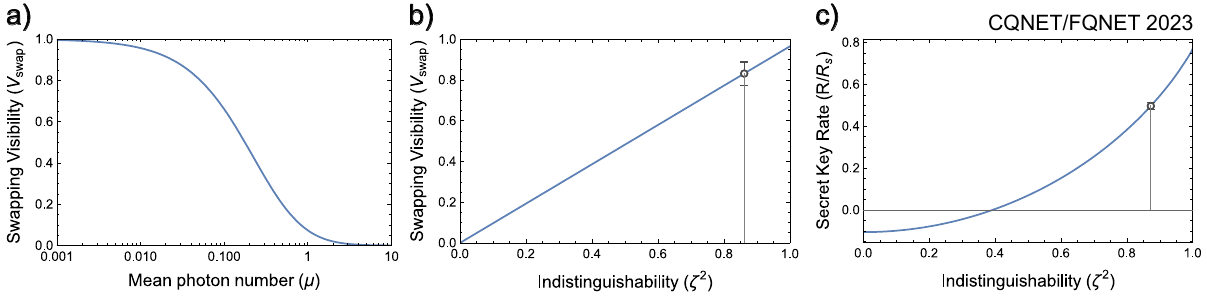}
  \caption{Theoretical models for teleportation of entanglement. a) Entanglement swapping visibility as a function of mean photon number for identical source mean photon numbers ($\mu = \mu_A = \mu_B$), unit path efficiencies, and unity indistinguishability. b) Entanglement swapping visibility as a function of indistinguishability for the experimentally characterized mean photon numbers and path efficiencies of the entanglement swapping measurements in Fig. \ref{fig:swapping}. The experimental swapping visibility is indicated with the circular marker. c) Secret key rate as a function of indistinguishability for the experimentally characterized mean photon numbers and path efficiencies of the QKD measurements in Table \ref{tab:qkd}. The experimental secret key rate is indicated with the circular marker.} 
  \label{fig:swapping_model}
\end{figure*}
\section{Entanglement swapping visibility}\label{sec:Vswap_app}

The entanglement swapping visibilities as a function of mean photon number and indistinguishability are shown in Fig. \ref{fig:swapping_model} a) and b), respectively. 
For the entanglement swapping measurement in Fig. \ref{fig:swapping}, the average swapping visibility of $\langle V_{\text{swap}} \rangle = 83.1\pm 5.5$\% corresponds to a photon indistinguishability of $\zeta^2 = 0.86\pm0.06$.
In Fig. \ref{fig:swapping_model}c, we plot the lower bound of Eq. (\ref{eq:secret_key_rate}) per sifted key rate as a function of indistinguishability for $\kappa = 1.22$, $e_t = 0.011$, and $e_p = (1-V_{\text{swap}})/2$, where $V_{\text{swap}}$ is the swapping visibility model for the QKD error rate measurement reported in Table \ref{tab:qkd}. The experimental secret key rate $ R/R_s = 0.50^{+0.18}_{-0.14}$ bits per sifted bit 
corresponds to an indistinguishability of $\zeta^2 = 0.87^{+0.09}_{-0.10}$. For completely indistinguishable photons and the same experimental parameters as the measurement in Fig. \ref{fig:swapping} (see Table. \ref{tab:exp_params}b), 
the model predicts an swapping visibility of $96.5\%$, corresponding to a swapping fidelity of $97.4\%$ and secret key rate of 0.87 bits per sifted bit.

\clearpage
\bibliography{references}







\end{document}